\documentclass[aps,twocolumn,secnumarabic,nobalancelastpage,amsmath,amssymb,nofootinbib]{revtex4}

\usepackage{subfigure}
\usepackage{graphicx}
\usepackage{wrapfig}
\usepackage{subfigure}
\usepackage{fancyhdr}
\usepackage{xfrac}
\usepackage{url}
\usepackage{ulem}
\usepackage{natbib}
\usepackage{appendix} 				
\usepackage{hyperref}				
\usepackage{chngcntr}				
\usepackage{everypage}				
\usepackage{pdflscape}
\usepackage{enumitem}				
\usepackage[english]{babel}
\fancypagestyle{plain}
\renewcommand{\figurename}{Figure}

\begin{document}

\title{Estimating sample-specific regulatory networks}
\author{Marieke Lydia Kuijjer\,$^{1,2,\dagger}$, Matthew Tung\,$^{1,2,\dagger}$\, GuoCheng Yuan$^{1,2}$, John Quackenbush$^{1,2,3}$, Kimberly Glass$^{4}$\footnote{contact: kglass@jimmy.harvard.edu}}
\affiliation{$^{1}$Department of Biostatistics and Computational Biology, Dana-Farber Cancer Institute, Boston, MA, USA\\
$^{2}$Department of Biostatistics, Harvard School of Public Health, Boston, MA, USA\\
$^{3}$Department of Cancer Biology, Dana-Farber Cancer Institute, Boston, MA, USA\\
$^{4}$Channing Division of Network Medicine, Department of Medicine, Brigham and Women's Hospital, Harvard Medical School, Boston, MA}

\begin{abstract}
Biological systems are driven by intricate interactions among the complex array of molecules that comprise the cell. Many methods have been developed to reconstruct network models of those interactions. These methods often draw on large numbers of samples with measured gene expression profiles to infer connections between genes (or gene products). The result is an aggregate network model representing a single estimate for the likelihood of each interaction, or ``edge,'' in the network. While informative, aggregate models fail to capture the heterogeneity that is represented in any population. Here we propose a method to reverse engineer {\it sample-specific networks} from aggregate network models. We demonstrate the accuracy and applicability of our approach in several data sets, including simulated data, microarray expression data from synchronized yeast cells, and RNA-seq data collected from human lymphoblastoid cell lines. We show that these sample-specific networks can be used to study changes in network topology across time and to characterize shifts in gene regulation that may not be apparent in expression data. We believe the ability to generate sample-specific networks will greatly facilitate the application of network methods to the increasingly large, complex, and heterogeneous multi-omic data sets that are currently being generated, and ultimately support the emerging field of precision network medicine.
\end{abstract}

\maketitle

\section{Introduction}\label{introduction}
In many instances, especially when analyzing complex traits and diseases, a single gene or pathway cannot fully explain a particular phenotype. In these cases, biological processes are often characterized as complex networks whose structures are altered as the phenotype changes. Studying the pattern of connections between biological components, and how these structures change between cell states, can yield new insights into the mechanisms driving disease. However, accurately reconstructing these networks in a way that captures both the properties and complexities of each phenotype remains a significant challenge.

Biological and phenotypic variability is a prominent feature in many complex traits and diseases. The generation of large multi-omic resources, including The Cancer Genome Atlas (TCGA), the ENCyclopedia Of DNA Elements (ENCODE)~\cite{encode2012integrated}, and the Genotype-Tissue Expression (GTEx)~\cite{gtex2015genotype,gtex2017genetic} project, as well as the recent rise of single-cell genomic technologies and the cataloguing of individual cell-types in the Human Cell Atlas~\cite{rozenblatt2017human}, have brought this issue to the forefront. We now recognize that diversity in the regulatory processes active in different cells, across multiple tissues, between various phenotypes, and even in response to environmental exposures, all contribute to the complexity of observed disease manifestations. It is also increasingly clear that the cumulative effect of multiple individual-specific variations, each with a relatively small effect-size, likely play an important role in the manifestation of many different diseases, including rare disease subtypes~\cite{mcclellan2010genetic}. These observations speak to a multi-factorial process. In other words, rather than individual molecules, it is alterations in biological processes, characterized as complex networks, that play a critical role in mediating the observed diversity~\cite{loscalzo2007human}. Effectively capturing this network-level heterogeneity is critical as we seek to understand how gene expression and regulatory processes manifest at an increasingly individualized level.

Existing methods for estimating biological networks often rely upon combining information from large quantities of data (most commonly gene expression data). This means that even when the data represents a spectrum of phenotypes, these approaches, by default, estimate only a single ``aggregate'' network~\cite{de2010advantages,marbach2012wisdom}. Although these types of aggregate networks have allowed us to gain important insights across a wide range of biological systems and diseases, they only capture the regulatory processes shared across a population of samples. More recently, several approaches have been suggested for exploring sample-level network information~\cite{alvarez2016functional,liu2015identification,liu2016personalized}. However, these methods are severely limited. In particular, current single-sample methods rely upon differential-analysis of the underlying expression data, thereby masking any information shared across the population (see section~\ref{supmeth_current} and Supplemental Table~\ref{suptab1}). Regulatory processes act on a network that contains both common and context-specific interactions~\cite{sonawane2017understanding}. However, there are currently no existing approaches designed to reconstruct the complete network for each sample in a population.

\begin{figure}[htbp]
  \includegraphics[width=250px]{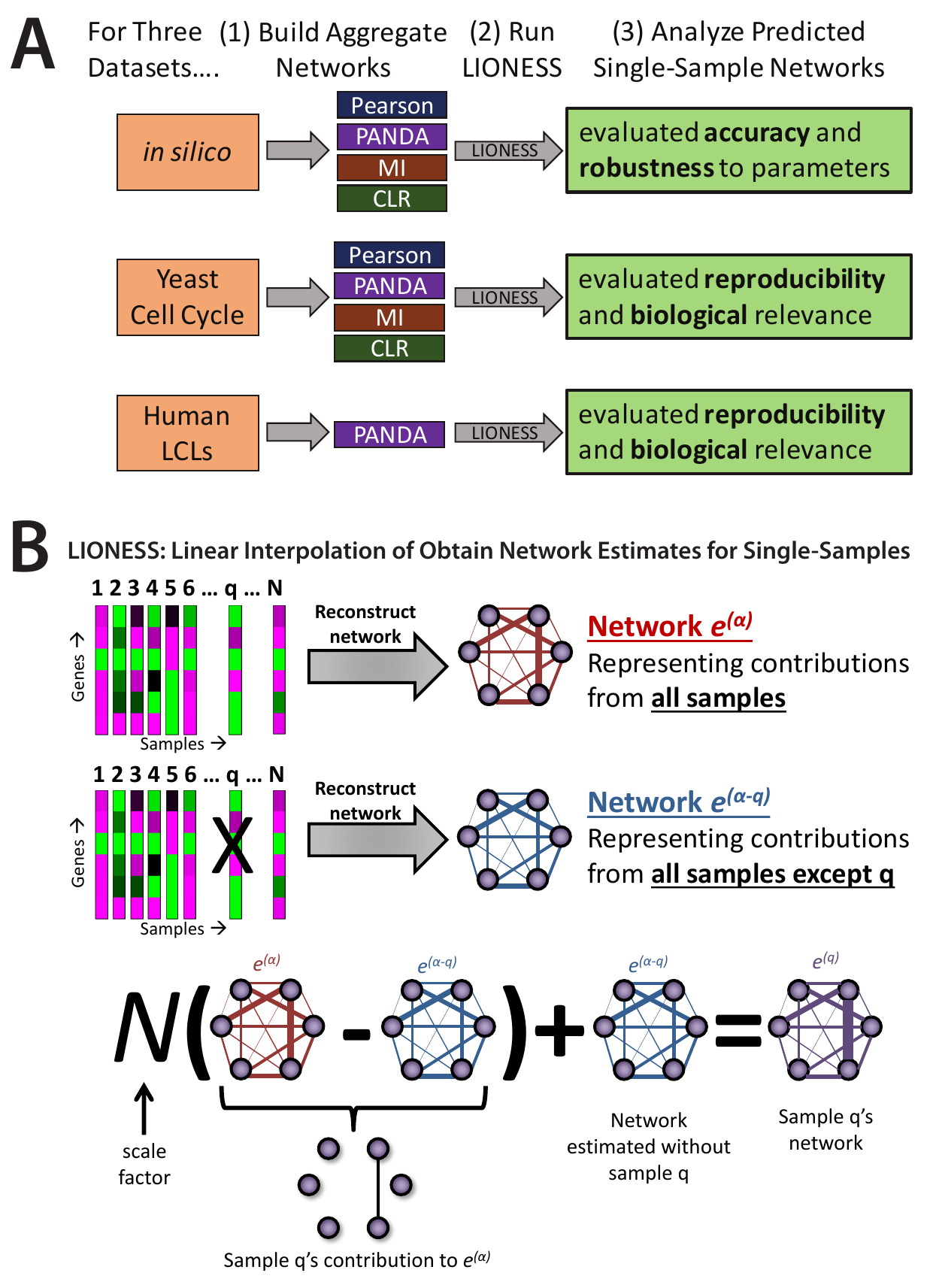}
  \caption{(A) Flow diagram summarizing of the analyses performed in this manuscript to evaluate the LIONESS approach. LIONESS was applied to multiple aggregate network reconstruction approaches including Pearson correlation coefficient, PANDA (Passing Attributes between Networks for Data Assimilation), MI (mutual information), and CLR (Context Likelihood of Relatedness). (B) Visual illustration of how LIONESS estimates the network for a single sample based on two aggregate network models, one reconstructed using all biological samples in a given data set, and one using all except the sample of interest ($q$).}
  \label{fig1}
\end{figure}

In order to fill this gap and effectively model the regulatory processes active in each sample in a population, we have developed a method to reverse engineer sample-specific networks. We call this approach LIONESS (Linear Interpolation to Obtain Network Estimates for Single Samples). LIONESS estimates individual sample networks by applying linear interpolation to the predictions made by existing aggregate network inference approaches. In this manuscript, we demonstrate the accuracy, robustness, and applicability of LIONESS in the context of multiple aggregate network reconstruction approaches and in several data sets, including simulated data, microarray expression data from synchronized yeast cells, and RNA-seq data collected from human lymphoblastoid cell lines (Figure~\ref{fig1}A; Supplemental Table~\ref{suptab2}). We also show how the predictions from LIONESS can be used to model regulatory network changes over time and to characterize the regulatory processes active in individual samples. Ultimately, we find analyzing single-sample regulatory networks provides a view of biological systems that is distinct from, but complementary to, other sources of multi-omic data.

\section{Methods}\label{methods}

\subsection{Complex relationships in biological networks}\label{meth_relationships}
Many widely-used network inference methods start by calculating a score or statistic for each gene pair based on shared information across a set of input gene expression samples~\cite{de2010advantages,marbach2012wisdom}. These scores are sometimes augmented to better account for regulatory complexity~\cite{faith2007large,margolin2006aracne,langfelder2008wgcna} but are ultimately used to infer the presence or absence of ``interactions'' between genes. This collection of genes and their corresponding complex set of inferred interactions are conceptualized as a network in which ``nodes'' represent genes and ``edges'' represent the interactions between those genes. In this context, heterogeneity in the underlying input samples is often essential for correctly estimating a network model, as variance in the data can amplify gene co-variation patterns, leading to more robust network predictions. However, at the same time, building this type of consensus, or ``aggregate,'' network model largely ignores the fact that there may be multiple different underlying regulatory networks represented across the individual input samples.

Consider the collection of cells within a tissue. We now recognize that within this system, each cell may have its own unique gene expression profile and corresponding unique active gene regulatory network. In the same way, each individual person in a group manifests a phenotype in a slightly different fashion, meaning that his or her gene expression profile and the gene regulatory network driving it should be subtly different. While we have started to embrace this complexity in analyzing gene expression, it has been largely ignored in the analysis of gene regulatory networks.

To better model network-level diversity across a population, we sought to develop a method that could model sample-specific networks. In developing our approach, we recognized that there are two types of relationships that needed to be considered: (1) intra-network relationships, or the connections among the nodes (genes) {\it within} a biological network, and (2) inter-network relationships, or the relationships {\it between} multiple different biological networks. The first of these (intra-network relationships) is an area that has been highly-studied. It is now widely recognized that relationships among nodes within a biological network are very complex and that these networks are often characterized by nonlinear regulatory dynamics and synergistic effects. Fortunately, there are many approaches that have already been developed to model these complex interactions~\cite{marbach2012wisdom,wang2014review}, as outlined above. In contrast, the comparative study of networks (inter-network relationships) is still a relatively young field. However, a number of recent studies have used linear approaches to analyze and cluster sets of networks~\cite{marbach2012wisdom,schlauch2017estimating,mucha2010community,onnela2012taxonomies}.

\subsection{LIONESS: Linear Interpolation to Obtain Network Estimates for Single Samples}\label{meth_lioness}
With the above in mind, we developed our approach by using a linear framework to relate a set of networks, each representing a different biological sample. In other words, we suggest that an ``aggregate'' network predicted from a set of $N$ samples can be thought of as the average of individual component networks reflecting the contributions from each member in the input sample set. Mathematically, this means that the weight of an edge, ${e}_{ij}^{({\alpha})}$ between two nodes ($i$ and $j$) in an aggregate network derived using all samples ($\alpha$) can be modeled as the linear combination of the weight of that edge across a set of networks:
\begin{equation}\label{main_Eq1}
{e}_{ij}^{(\alpha)}=\displaystyle\sum_{s=1}^{N} w_s^{(\alpha)} e_{ij}^{(s)}, \text{  where  } \displaystyle\sum_{s=1}^{N} w_s^{(\alpha)} = 1.
\end{equation}
In this equation, each network (${e}_{ij}^{(s)}$) in the set directly corresponds to one of the samples ($s$) used to reconstruct the aggregate network (${e}_{ij}^{({\alpha})}$), and $w_s^{(\alpha)}$ represents the relative contribution of that sample to the aggregate model; we note that the complex relationships between the nodes in the aggregate network (${e}_{ij}^{({\alpha})}$) can be calculated using any aggregate network reconstruction approach. This allows us to ensure that higher-order, nonlinear relationships, such as those commonly found in complex biological networks, can be included in the network models.

Next, we also suggest that, as in Equation~\ref{main_Eq1}, the weight of an edge in a network reconstructed using all but one of the samples ($\alpha-q$), can be written as:
\begin{equation}\label{main_Eq2}
{e}_{ij}^{(\alpha-q)}=\displaystyle\sum_{s\neq q}^{N} w_s^{(\alpha-q)} e_{ij}^{(s)}, \text{  where  } \displaystyle\sum_{s\neq q}^{N} w_s^{(\alpha-q)} = 1. 
\end{equation}
Comparing Equations~\ref{main_Eq1} and~\ref{main_Eq2}, we find that $w_q^{(\alpha)}=1-w_s^{(\alpha)}/w_s^{(\alpha-q)}$. This comparison also allows us to solve exactly for the network for an individual sample $q$. In particular, by subtracting the above equations we find:
%
\begin{align}
\label{main_Eq3}
{e}_{ij}^{(\alpha)}-{e}_{ij}^{(\alpha-q)}&={w}_{q}^{(\alpha)}{e}_{ij}^{(q)}+\displaystyle\sum_{s\neq q}^{N}(w_s^{(\alpha)}-w_s^{(\alpha-q)}){e}_{ij}^{(s)}\\
&={w}_{q}^{(\alpha)}{e}_{ij}^{(q)}-{w}_{q}^{(\alpha)}\displaystyle\sum_{s\neq q}^{N}{w}_{s}^{(\alpha-q)}{e}_{ij}^{(s)}.
\end{align}
%
The network specific to sample $q$ in terms of the aggregate networks is then:
\begin{equation}\label{main_Eq4}
{e}_{ij}^{(q)}=\frac{1}{w_q^{(\alpha)}}(e_{ij}^{(\alpha)}-e_{ij}^{(\alpha-q)})+e_{ij}^{(\alpha-q)}.
\end{equation}

In summary, the edge scores for a given individual network are equal to the difference in edge scores for an aggregate network constructed using all the samples and an aggregate network reconstructed using all but the sample of interest, multiplied by a scaling factor, and added to the edge scores of the network reconstructed using all but the sample of interest (Figure~\ref{fig1}B). What this means is that we can use pairs of aggregate network models to ``extract'' networks for each of the individual input samples. In the following analysis we give samples equal weight (${w}_{q}^{(\alpha)}={1}/{N}$) although one could, in principle, weight samples differently based on the quality of the data for individual samples or some other measure. A more detailed version of the LIONESS derivation is provided in section~\ref{SuppMat}.

We note that the mathematical framework presented in Equation~\ref{main_Eq4} is independent of the inference method used to estimate the aggregate network edge-weights. In other words, LIONESS can be thought of as a mathematical ``wrapper'' that can be applied to estimate networks based on any aggregation model. With this in mind, we have performed a detailed exploration of the behavior of Equation~\ref{main_Eq4} when the aggregate network model is calculated using Pearson correlation or mutual information, two measures commonly applied to quantify the level of a linear or nonlinear association between variables, respectively. For both measures, we are able to show the inter-network linearity assumption of LIONESS (Equation~\ref{main_Eq1}) holds in the context of large sample size (see section~\ref{SuppMat}). Simulation analysis also illustrates how LIONESS consistently assigns similar edge-weights to the samples that most contribute to an expected relationship, and correctly identifies and re-weights edges for the samples that are most inconsistent with an expected relationship (see section~\ref{supmeth_intuition} and Supplemental Figure~\ref{supfig1}).

\section{Results}\label{results}

\subsection{LIONESS accurately and reproducibly predicts networks using {\it in silico} data}\label{insilico}
To systematically evaluate LIONESS, we created a series of data sets where the underlying networks corresponding to each input expression sample are known. We used these data to (1) evaluate whether LIONESS accurately predicts individual sample networks, (2) to explore how sensitive these predictions are to the properties of the underlying data, and (3) to assess whether LIONESS is able to recover sample-specific network relationships (i.e. edges specific to a given sample's network).

\begin{figure*}[htbp]
  \includegraphics[width=500px]{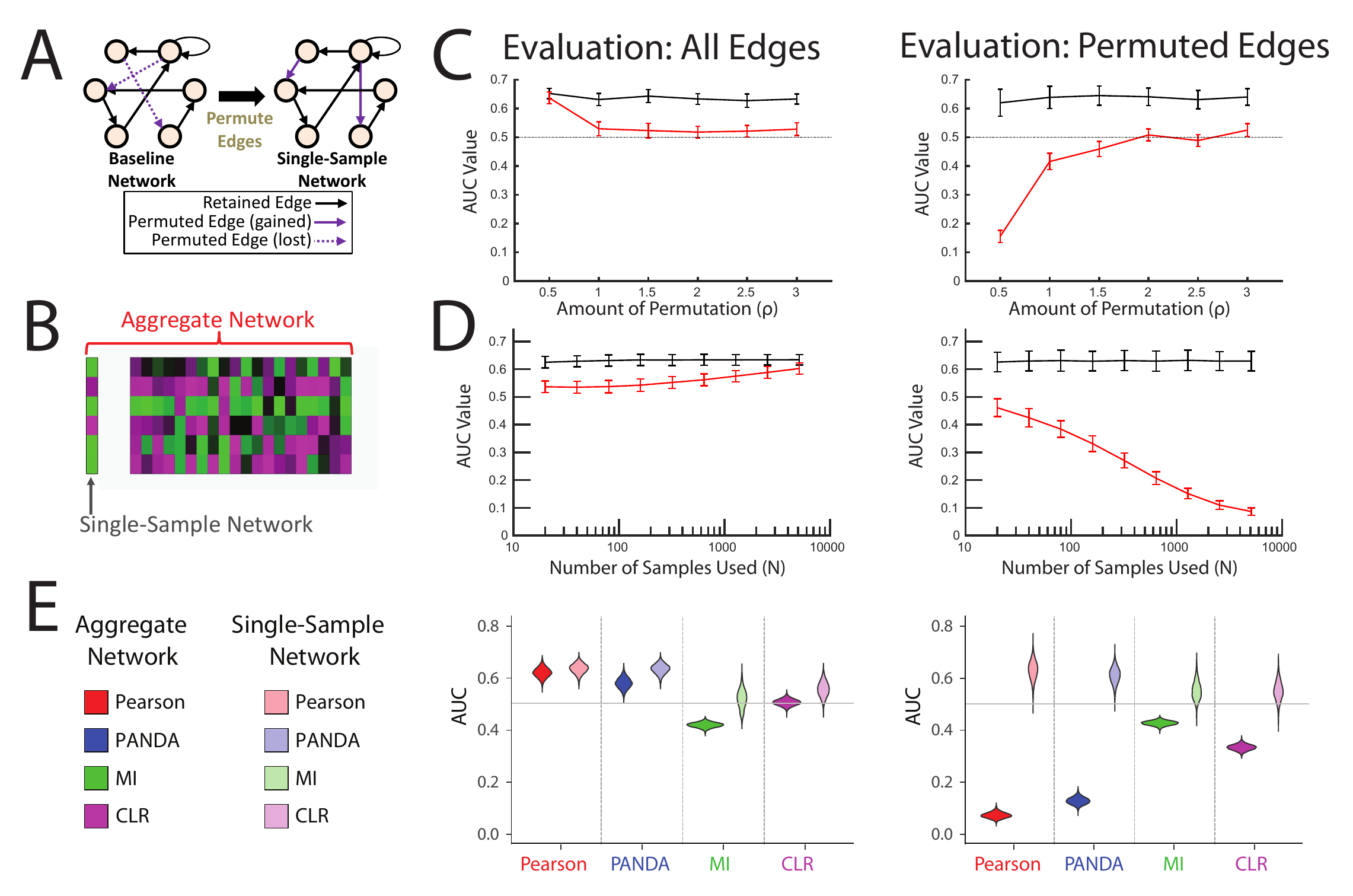}
  \caption{Evaluation of LIONESS' ability to recover known single-samples networks in {\it in silico} data. (A) Toy example of how we create a single-sample network from an underlying baseline network. (B) Illustration of the gene expression samples used to build a single-sample network. We evaluated the accuracy of both the aggregate network derived using all samples (red) and the LIONESS-estimated single-sample network (black) by benchmarking against the corresponding ``gold-standard'' single-sample network. (C) The mean and standard deviation of the AUC values of the aggregate (red) and LIONESS-predicted single-sample networks (black) estimated from {\it in silico} data sets representing varying levels of heterogeneity. (D) The mean and standard deviation of the AUC values of the aggregate (red) and LIONESS-predicted single-sample networks (black) estimated using increasing numbers of input expression samples. For each sample size, 10000 random subsets of samples were used. (E) Violin plots showing the distribution of AUC values for aggregate and LIONESS-predicted single-sample networks estimated using four different aggregate network reconstruction approaches. For (C)--(E) AUCs were calculated both using all possible edges, and for edges that differ from the baseline model (permuted edges), see (A).}
  \label{fig2}
\end{figure*}

Briefly, to create a benchmark {\it in silico} data set, we started with a baseline network containing $M$ nodes and random edges. We then permuted the edges within this baseline network, creating a single-sample network with the same degree distribution (Figure~\ref{fig2}A). We repeated this $N$ times, creating $N$ ``gold standard'' single-sample networks. To derive corresponding expression profiles for each of these networks, we generated 1000 random initial expression states (0 or 1 corresponding to whether the gene is ``on'' or ``off'') and applied a Boolean model (see section~\ref{supmeth_data}) to determine the corresponding network attractors~\cite{wuensche1998discrete}. We averaged over all states defined within these 1000 attractors to generate ``expression'' values for the $M$ nodes (which represent genes) in each network. This gave us an $M$-by-$N$ matrix of expression values, one for each of the nodes (genes) in each network. An overview of our approach is shown in Supplemental Figure~\ref{supfig2}.

We first evaluated LIONESS' predictions in the context of varying heterogeneity. To do this, we generated six different {\it in silico} data sets using the same baseline network but varying the amount of permutation used to obtain the single-sample network models. For this analysis we chose a network size of $M=100$ nodes and $N=100$ samples and used Pearson correlation to calculate an aggregate network before applying Equation~\ref{main_Eq4} to reconstruct each of the individual sample networks. We evaluated the accuracy of the Pearson correlation aggregate network and each of the LIONESS-estimated single-sample networks (Figure~\ref{fig2}B) by comparing with the original ``gold standard'' networks and calculating the Area Under the Receiver Operator Characteristic curve (AUCROC, or more simply AUC).

We observe that in the context of greater heterogeneity among the single-sample networks (increased permutation) the LIONESS-predicted networks are much more accurate than the aggregate network (Figure~\ref{fig2}C). On the other hand, in the context of low heterogeneity, the accuracy of the LIONESS-predicted networks is similar to that of the aggregate network; this is to be expected since the aggregate network should not be significantly different from the single-sample networks in this context. Most interesting, however, is the fact that the accuracy of the permuted edges (those that appear in the single-sample network but not the baseline network, see Figure 2A) is {\it independent} of sample heterogeneity. These edges are not accurately captured in the aggregate network model, especially in the case of low-heterogeneity.

We have repeated this analysis on {\it in silico} data for networks (1) of various sizes (contain more nodes) and (2) with varying levels of noise added to their associated expression data. We find that LIONESS' performance is independent of the size of the network models (Supplemental Figure~\ref{supfig3}A--B), and retains its ability to predict networks even in the presence of expression data noise (Supplemental Figure~\ref{supfig3}C).

Next, we evaluated LIONESS' predictions in the context of varying sample size. To do this, we generated an additional {\it in silico} data based on the same 100-node baseline network as the previous analysis. We used a moderate level of permutation ($\rho=1$) to generate a data set with ten thousand paired network and expression samples. We selected subsets of these data containing $N+1$ samples, where $N$ varied from 20 to 5000, applied LIONESS to estimate the $(N+1)^{th}$­ sample's network, and evaluated the accuracy of that network as well as the corresponding aggregate network from which it was derived (Figure~\ref{fig2}D). We observe that as we increase the number of samples ($N$), the accuracy of LIONESS single-sample networks remains constant, both overall and for the sample-specific permuted edges. However, although including more samples improves the accuracy of the aggregate network model, the sample-specific permuted edges within the aggregate model are very poorly estimated with increasing sample-size. This behavior is expected; including more samples provides increasing information that can help accurately estimate edges that are in the baseline network (those that are most likely to be common across all the single-sample networks). These edges are---by definition---the opposite of the sample-specific permuted edges.

Finally, we tested the generalizability of LIONESS by estimating single-sample networks from aggregate models derived using several common network reconstruction approaches, including Pearson correlation, PANDA~\cite{glass2013passing}, mutual information, and CLR~\cite{faith2007large} (for more information, see section~\ref{supmeth_networks}). Figure~\ref{fig2}E shows the distribution in AUC values for the aggregate and LIONESS single-sample network predictions for each of these approaches. We find that LIONESS consistently and accurately predicts single-sample networks for all four network inference methods. Interestingly, although the difference in AUC between the overall aggregate and single-sample models is fairly similar for all four approaches, the AUC values are lowest for networks estimated using mutual information, a nonlinear approach for assessing correlation. This may reflect that our {\it in silico} data doesn't fully represent the complexity found in biological systems or that mutual information is not the optimal measure to use when estimating a regulatory network from expression data.

\subsection{Estimating single-sample networks using experimental data from yeast}\label{yeast}
We next tested LIONESS using experimental data from cell-cycle synchronized yeast cells. We downloaded gene expression data (GEO accession, GSE4987)~\cite{pramila2006forkhead} consisting of dye-swap technical replicates measured every five minutes for 120 minutes. We ma-normalized~\cite{yang2007marray} these data, removed probe sets with missing information, batch-corrected using ComBat~\cite{johnson2007adjusting}, averaged probe sets mapping to the same ORF annotation, and quantile-normalized the resulting gene-by-sample matrix of expression values. We note that the 105 minute time point was excluded in both replicates due to poor hybridization performance~\cite{pramila2006forkhead}.

\begin{figure}[htbp]
  \includegraphics[width=250px]{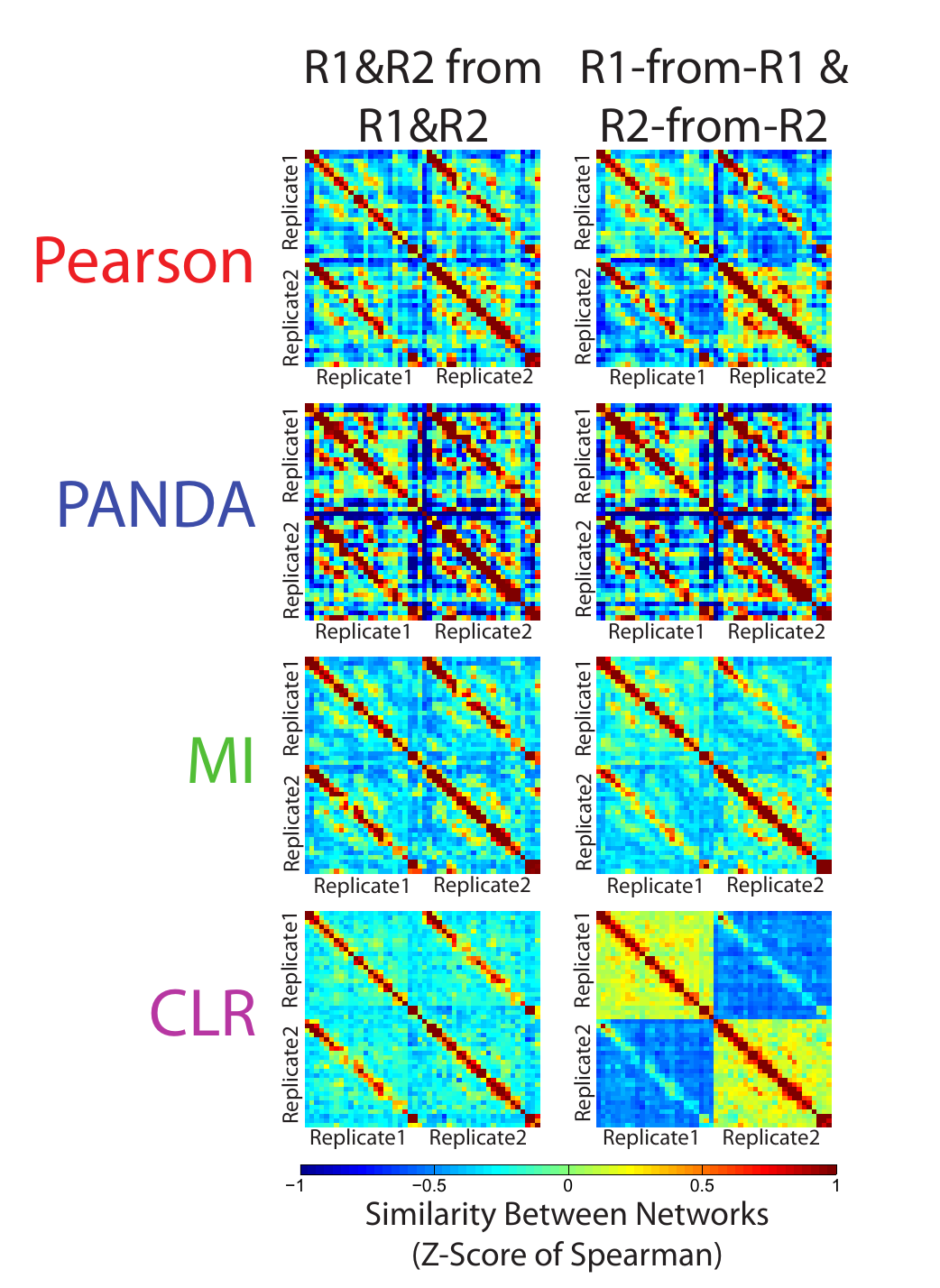}
  \caption{Analysis of LIONESS networks predicted for 48 expression samples collected across a yeast cell-cycle time course experiment. LIONESS was used to predict networks for each sample in the expression data by applying four different aggregate network reconstruction approaches. For each approach we built the aggregate models either using all samples (R1\&R2 from R1\&R2), or only the samples from the same technical replicate (R1-from-R1 \& R2-from-R2). The Spearman correlation was used to evaluate how similar these networks are to each other.}
  \label{fig3}
\end{figure}

We used four different network inference methods (Pearson Correlation, PANDA~\cite{glass2013passing}, mutual information, and CLR~\cite{faith2007large}) to reconstruct aggregate networks for this data set and applied LIONESS to estimate the networks for each of the individual samples. The correlation between edge weights in each pair of the estimated sample-specific networks is shown in the first column of Figure~\ref{fig3} (R1\&R2-from-R1\&R2). We see that network estimates for the same technical replicate are highly similar, as evidenced by the strong diagonal in the upper-right and lower-left square of each comparison; additional structure is also evident in off-diagonal similarities that reflect the fact that the time course data includes more than one cell cycle.

To test if strong reproducibility was due to including replicates in the expression data, we also ran LIONESS separately on each individual replicate. This analysis produced 24 single-sample networks estimated using only the data in replicate one, and 24 single-sample networks estimated using only the data in replicate two (R1-from-R1 \& R2-from-R2). The correlation between these networks is shown in the second column of Figure~\ref{fig3}. As before, we observe strong reproducibility in estimated edge weights between technical replicates. However, it is worth noting that even though we have corrected for batch effects in the expression data, several of the methods, especially CLR, appear to be sensitive to the ``background'' data used.

We note that this level of reproducibility is similar to that observed in the underlying expression data, demonstrating that we did not lose replicate information by applying LIONESS separately to the two sets of expression samples (Supplemental Figure~\ref{supfig4}A). Interestingly, replicate PANDA networks had higher levels of similarity as compared to the other three reconstruction approaches. Based on these results, in the following analysis we focus on the single-sample networks derived using PANDA as the aggregate network inference method. Results for the other reconstruction approaches are presented in Supplemental Figure~\ref{supfig4}B.

\subsection{Single-sample networks show periodic structure across the cell cycle}\label{cellcycle}
\begin{figure*}[htbp]
  \includegraphics[width=500px]{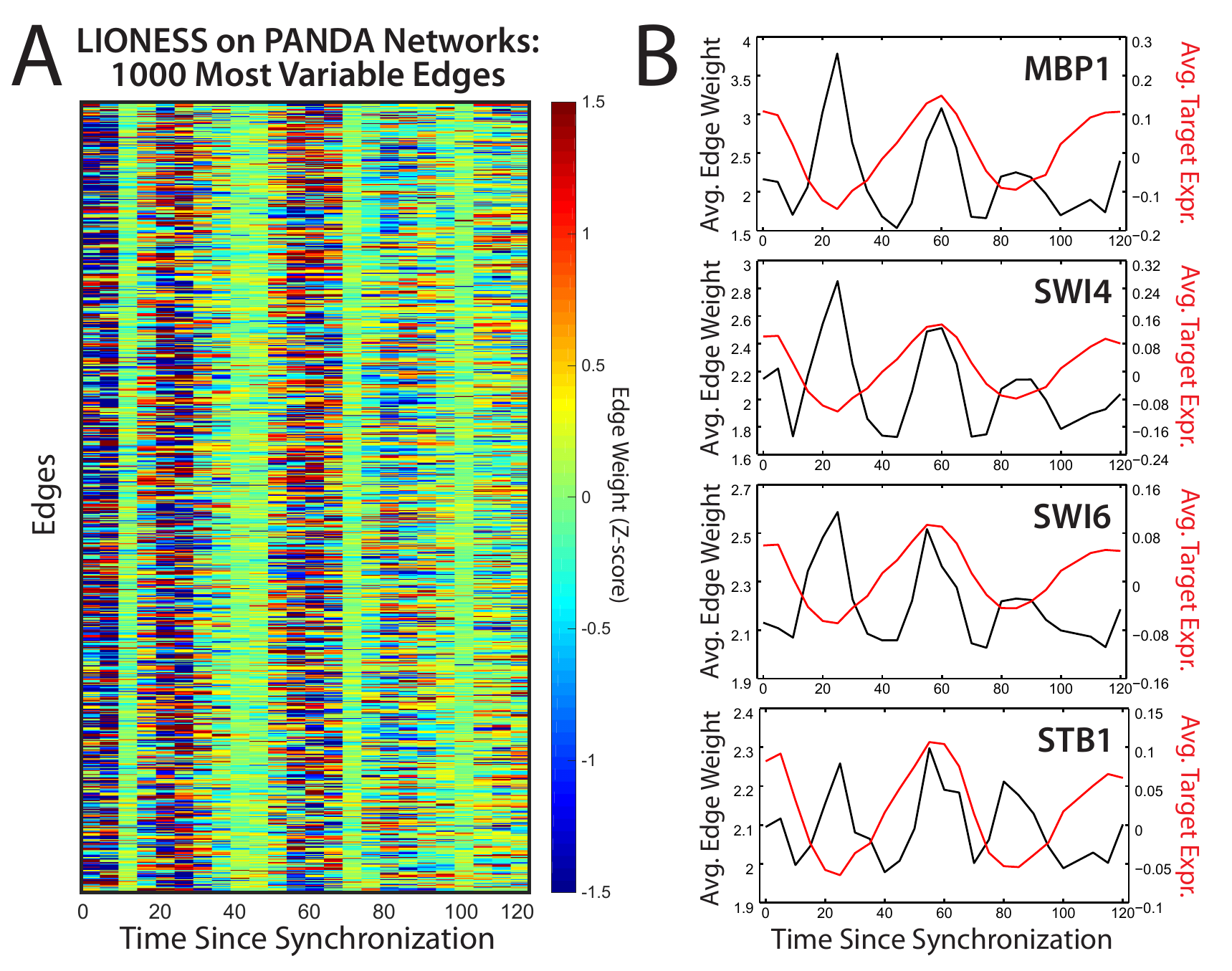}
  \caption{Characterizing networks across the yeast cell-cycle. (A) A heat map of the edge weights for the 1000 most variable edges across the sample-specific network models. Rows are Z-score normalized for visualization purposes. (B) The average expression of genes targeted by the four transcription factors that were identified as regulatory nodes of the 1000 top most variable edges as well as the average weight of high-confidence edges that extend between those transcription factors and their target genes.}
  \label{fig4}
\end{figure*}

We next tested whether these single-sample networks could provide insight into gene regulation and dynamic cellular network processes. We averaged sample networks representing the same time point in each of the two replicates, identified the 1000 edges with the highest variability across the individual networks, and visualize those edges as a heat map in Figure~\ref{fig4}A. We observe strong oscillatory patterns in edge weights, apparently reflecting changes in gene regulation across the cell cycle. Further investigation indicates that all these highly variable edges originate from one of four transcription factors (MBP1, SWI4, SWI6, and STB1), each of which is known to play a key role in regulating the yeast cell cycle~\cite{ho1999regulation}.

We examined the genes for which there is strong evidence of targeting by these transcription factors (average edge weight across all LIONESS networks greater than zero). In Figure~\ref{fig4}B we plot the average weight of these high-evidence interactions for each regulating transcription factor and the average expression of their target genes. It is immediately apparent that oscillation in edge weights occurs at exactly twice the frequency of the oscillation in gene expression, and that the gene expression oscillates with a period approximately equal to that of the yeast cell cycle.

To understand this result we have to recognize that PANDA interprets correlation in target gene expression as an indication of co-regulation by an upstream transcription factor. Consequently, PANDA assigns greater edge weights when a transcription factor's targets are all coordinately increasing (activated) or decreasing (de-activated or repressed) their expression levels. High edge weights should be interpreted as evidence for information flow from a transcription factor (TF) to its targets, which could be due to a physically present TF actively regulating its downstream targets, but could also reflect a strong {\it lack} of regulation by that TF. In this light, the ``turn on/turn off'' behavior is exactly what one would predict given how PANDA estimates network relationships and is further evidence that LIONESS is extracting meaningful single-sample networks.

\subsection{Reconstructing single-sample networks for human lymphoblastoid cell lines}\label{human}
Lastly, we applied LIONESS to infer individual-specific human gene regulatory networks. We used a set of 155 RNA-seq samples from immortalized lymphoblastoid cell lines representing 65 different individuals~\cite{pickrell2010understanding}. We downloaded raw fastq files from the Pritchard lab website (\url{http://eqtl.uchicago.edu/}) and aligned samples to hg19 using Bowtie~\cite{langmead2009ultrafast}; subsequent quality control analysis using RNA-SeQC~\cite{deluca2012rna} excluded two samples due to low expression profile efficiency scores. This left us with a final set of 153 RNA-seq experiments that includes replicates and represents 65 distinct individuals. We normalized these data using DEseq2~\cite{love2014moderated}. For additional data processing and normalization information, see section~\ref{supmeth_data}.

Based on our results when applying LIONESS to network models in the simulated and yeast cell cycle data, we chose PANDA as our aggregate network reconstruction method for the human data. We used PANDA to estimate aggregate gene regulatory network models for the collection of 153 RNA-seq samples. We then applied LIONESS to these aggregate models, resulting in 153 single-sample networks, one for each of the RNA-seq expression samples. A hierarchical clustering (complete linkage, Spearman Correlation) of the network edge weights demonstrates that networks for the same individual nearly always cluster more strongly with each other than with networks representing different individuals (Supplemental Figure~\ref{supfig5}). This analysis demonstrates that even when constructing networks using biological data from higher-order organisms such as human, the sample-specific networks predicted by LIONESS are reproducible.

\subsection{Complex relationships between network targeting and gene expression}\label{targeting}
As with yeast, we investigated the relationship between gene targeting and expression in human networks. First, we averaged single-sample networks that represent the same individual, resulting in 65 ``person-specific'' regulatory networks. We then selected high-evidence regulatory interactions for each transcription factor (average edge-weight across all single-sample networks greater than zero), and directly compared the mean edge-weight for these interactions in each of the single-sample networks to the average expression of the targeted genes in the original expression samples.

\begin{figure*}[htbp]
  \includegraphics[width=500px]{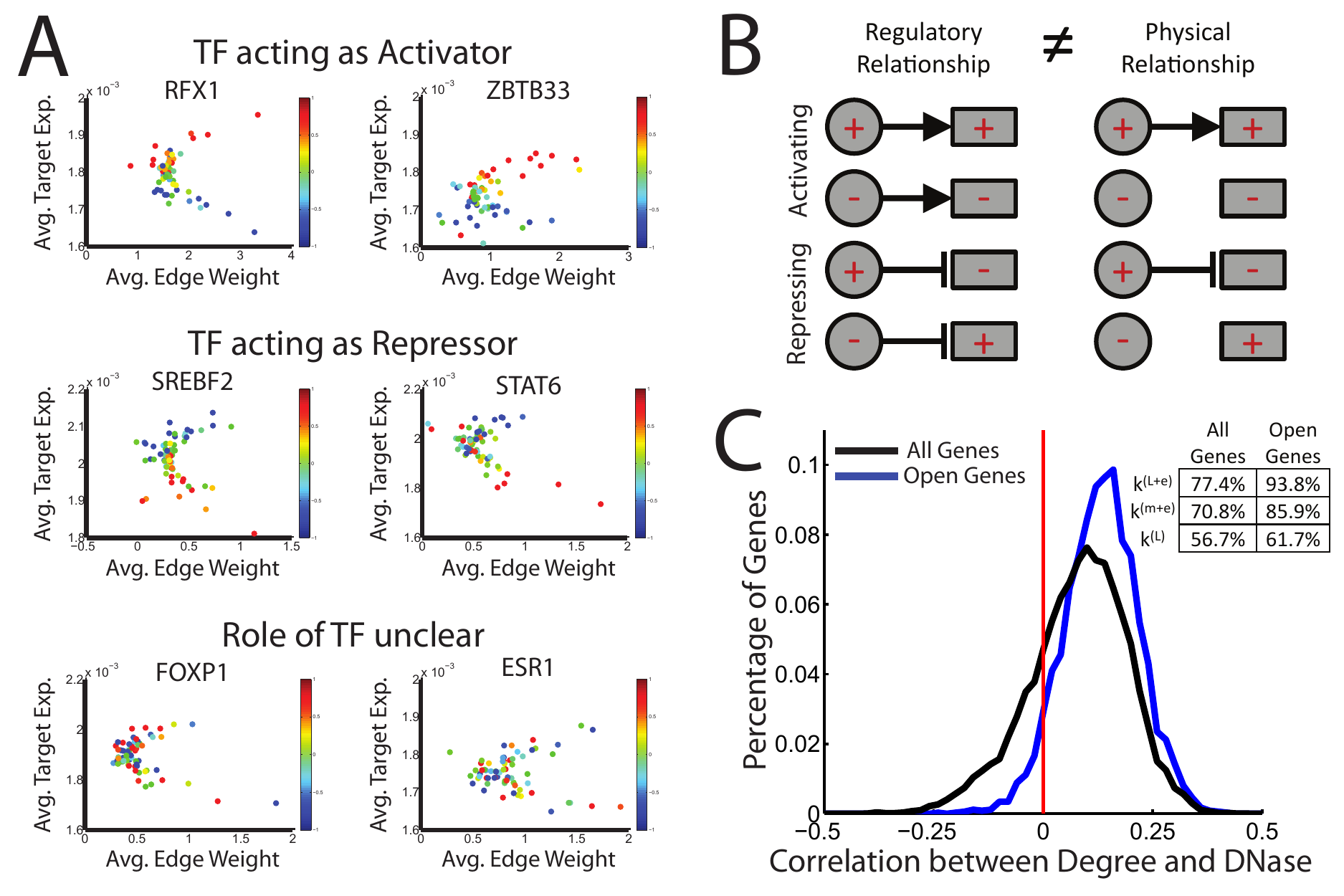}
  \caption{Comparison of gene regulation, gene expression and DNase hypersensitivity data. (A) For six representative transcription factors, the mean expression of target genes and the mean weight of the edges targeting those genes across the 65 samples are plotted. For each sample, the expression of the TF is shown as a color, scaled to the normal distribution for visualization purposes. (B) A cartoon illustrating how high edge weights and thus regulatory activity is not necessarily equivalent to the presence of a physical interaction. (C) The distribution of the Spearman correlation values when comparing gene-targeting (calculated by combining LIONESS predictions with TF expression; $k^{(L+e)}$) and the significance level of DNase hypersensitivity in a gene's promoter across all the samples. We also show the percentage of genes whose targeting positively correlates with DNase hypersensitivity when targeting is calculated using only the LIONESS-predicted edge weight ($k^{(L)}$: no expression considered) or a combination of expression and motif information ($k^{(m+e)}$). We performed these analyses either using all 12424 genes included in our network model, or for the set of 3488 genes with a DNase-peak called in all 65 samples (open genes).}
  \label{fig5}
\end{figure*}

We found nonlinear relationships between targeting and expression, with the highest average edge weights occurring when target genes have either high or low expression levels (Figure~\ref{fig5}A); this is consistent with what we observed in our yeast analysis (Figure~\ref{fig4}B). Coloring by the transcription factor expression level in each sample reveals additional patterns with some transcription factors primarily acting as activators (increased target gene expression upon increased TF expression and targeting) and others generally acting as repressors (decreased target gene expression upon increased TF expression and targeting). However, the relationship between a transcription factor and its target genes is not always simple, indicating that other regulatory mechanisms, such as co-activators, post-translational modifiers, or epigenetic mechanisms, are likely playing an important role in mediating these regulatory events.

\subsection{Increased network targeting corresponds to open chromatin}\label{openchromatin}
DNase hypersensitivity profiling data is also available for these 65 lymphoblastoid cell lines~\cite{degner2012dnase}, and we used it to investigate how network structures reflect epigenetic state. We downloaded the data from the Pritchard lab website (\url{http://eqtl.uchicago.edu/}) and called DNase ``peaks'' for each sample using MACS~\cite{zhang2008model}. When a peak fell within the promoter region of a gene, we assigned that gene a sample-specific score reflecting the significance level of the associated peak call. We found 12424 genes with a DNase promoter-peak in at least one sample and 3488 with a promoter-peak in all samples. For details on the DNase data processing, see section~\ref{supmeth_data}.

A DNase hypersensitivity peak represents a region of open chromatin that is often presumed to be occupied by one or more regulatory proteins, including transcription factors. We wanted to determine if differences in chromatin state between the 65 individuals is reflected in alterations in transcription factor targeting within our single-sample networks. We assigned each edge in each sample a score by combining (1) the weight of that interaction in our single-sample network models (since this value indicates whether information is flowing between that transcription factor and target gene in the PANDA model) and (2) the expression level of the transcription factor itself (since this value indicates whether the TF is physically present in the cell (Figure~\ref{fig5}B)). This resulted in a set of expression-modified edge-weights for each sample. For more information on how we calculated these edge-weights, see section~\ref{supmeth_analysis}.

We next used the sum of the edge-weights associated with each gene to estimate the number of transcription factors regulating that gene in each of the 65 person-specific networks ($k^{(L+e)}$). For comparison, we calculated gene-targeting two other ways: (1) using LIONESS edge-weight estimates in the absence of gene expression information ($k^{(L)}$) and (2) using gene expression information in the absence of LIONESS-predicted edge-weights ($k^{(m+e)}$); for the second measure we combined transcription factor expression in each sample with the motif information used for PANDA's prior (see section~\ref{supmeth_analysis}). We note this last approach is conceptually similar to current methods for approximating sample-specific network information (see Introduction, section~\ref{introduction}).

To evaluate the association of network targeting with chromatin state, for each gene we calculated the Spearman correlation between gene-targeting across the networks and the significance scores of that gene's promoter-DNase across the corresponding cell lines. We find that gene-targeting in the expression-modified LIONESS model ($k^{(L+e)}$) is very strongly correlated with promoter-DNase events, especially when only considering genes with measured chromatin information across all the cell lines (Figure~\ref{fig5}C). This association is greater than when using only expression and motif information ($k^{(m+e)}$), demonstrating that the LIONESS approach provides additional information on chromatin state not apparent in the data used to seed the algorithm.

\subsection{Differential-targeting of genes highlights important biological processes}\label{biology}
\begin{figure*}[htbp]
  \includegraphics[width=500px]{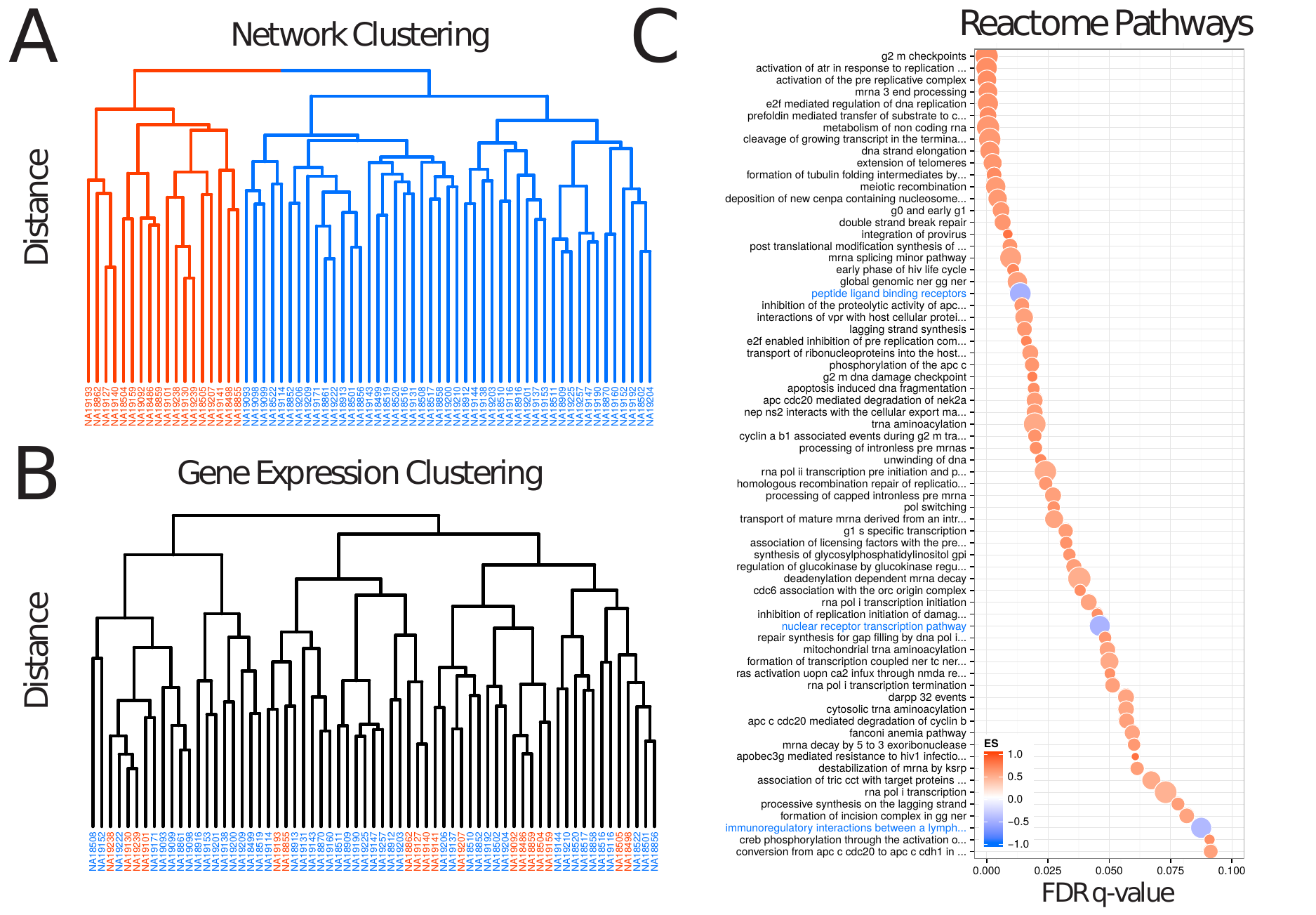}
  \caption{(A) A hierarchical clustering on the edge-weights for 65 regulatory networks, one for each subject included in the RNA-seq data set. (B) An equivalent hierarchical clustering on gene expression values for these 65 individuals. This clustering is distinct from the one based on the network edge weights. Subject labels are colored based on the network clustering. (C) Reactome pathways enriched based on GSEA using gene-targeting instead of gene expression and comparing samples from the right and left groups of networks presented in (A). No Reactome pathways were identified when comparing the expression values of genes in the different groups defined by the hierarchical clustering presented in (B).}
  \label{fig6}
\end{figure*}

Finally, we wanted to determine if there are common structures across these single-sample regulatory networks that might be reflective of important biological processes. We performed a hierarchical clustering (complete linkage, Spearman Correlation) on the edge-weights in the 65 single-sample networks and identified two distinct groups of samples defined by sample-specific edge weights (Figure~\ref{fig6}A). In parallel, we performed a hierarchical clustering using gene expression values (Figure~\ref{fig6}B) and found two groups of samples that are distinct from the groups defined by the edge-weight clustering.

We then used Gene Set Enrichment Analysis (GSEA)~\cite{subramanian2005gene} to compare groups of samples defined in the expression-based and network-based clustering. First, we compared the expression levels of genes between groups of individuals defined in the expression-based clustering. Although there were many differentially-expressed genes between the expression-based groups of samples (2620 with FDR $<0.01$), GSEA found no enrichment for known biological functions. Next, we defined the targeting-level of a gene in a sample as the sum of all edges pointing to that gene in a single-sample network. We then used GSEA to compare the targeting-levels of genes between the two groups of individuals defined in the network-based clustering~\cite{glass2014sexually}. In contrast to the expression-based analysis, in the differential-targeting analysis GSEA found enrichment for many cellular processes related to cell proliferation (in the smaller ``orange'' cluster; n $=18$) and immune function (in the larger ``blue'' cluster; n $=45$; Figure~\ref{fig6}C).

Unfortunately, there is little phenotypic information for the 65 individuals in this study, and those available~\cite{choy2008genetic} are not significantly associated with the groups defined by clustering on either the expression or the network information. However, given our functional enrichment results, we believe that the regulatory differences we observe between the network groups is likely related to differences in cellular growth rate induced by variable Epstein-Barr Virus (EBV) levels in the cell lines. EBV is used to transform human B-cells into immortalized lymphoblastoids and is known to activate NF-$\kappa$B transcriptional response~\cite{cahir1999epstein}. Consistent with this hypothesis, we find the signature ``Activation of NF-$\kappa$B in B-cells'' highly targeted in the small, ``cell proliferation'' cluster (ES $=0.5$, FDR $=2\cdot10^{-3}$).

Overall, these results indicate that evaluating single-sample networks can lend insight into the biological processes active in different individuals even when a similar analysis of the gene expression data does not.

\section{Discussion}\label{discussion}
In this paper, we present LIONESS as a method for estimating sample-specific regulatory networks. The core principle behind LIONESS is that the addition or removal of even a single sample will slightly perturb an aggregate network model. This perturbation can be used to estimate the contribution of a sample to the aggregate network, and therefore the network of that sample. Importantly, by relying on independent and existing aggregate models to capture the network of the interactions between genes, and a linear interpolation to estimate individual-level differences in the associated edge weights, LIONESS is able to reconstruct network estimates for each sample while preserving the biological complexity of the gene-gene interactions.

There are many network reconstruction methods but no consensus as to the ``correct'' or ``best'' one to use---if in fact there is a single method that works best for all data types~\cite{marbach2012wisdom}. In the analysis presented here, we used four representative gene expression network reconstruction approaches: Pearson correlation, Mutual Information, Context Likelihood of Relatedness (CLR), and PANDA. These were chosen because they illustrate network reconstruction methods that use either a linear (Pearson) or nonlinear (mutual information) correlation measure, and the extensions of those measures to better capture true regulatory interactions instead of simple correlative effects. Within this representative collection of methods, our analysis suggests that applying LIONESS to aggregate networks reconstructed using PANDA has the greatest potential for reconstructing accurate network models that can be used to interpret phenotypic differences.

We also note that although we tested our approach in the context of using gene expression to reverse-engineer regulatory networks, the linear algebraic framework at the heart of LIONESS is generalizable and can be applied in other settings where aggregate relationships are inferred from a collection of samples. In principle, this not only includes the application of LIONESS to other network inference methods, but also in contexts where network relationships are inferred from other multi-sample ‘omics data, such as metabolomics data, genetic/variant data, or epigenetic regulatory information such as CpG methylation. Incorporating this information into single-sample network models will be an important step in understanding the complexity of metazoan gene regulation.

Looking forward, LIONESS provides a way to unite the extensive literature and methodologies for modelling complex network relationships, with statistical analysis techniques that use sample-level information to model heterogeneity. Great progress has been made in assigning patients to disease subgroups based on gene expression profiles, or in using mutational profiles to match individual patients to specific therapies. LIONESS provides a framework in which one could imagine using a similar approach to analyze networks for precision medicine applications. For example, the network-interactions and properties predicted using LIONESS could be directly associated with patient phenotype, genotype, progression, survival, drug response, etc. Therefore, LIONESS not only addresses the problem of estimating multiple networks for populations with significant phenotypic of biological heterogeneity, it also provides a means of estimating and analyzing networks when samples of a particular phenotype or disease subtype are rare. Ultimately, one could imagine using the LIONESS approach to identify and target the regulatory pathways active in an individual patient (rather than using mutations or gene expression as surrogates for those pathways).

In summary, our approach to modeling single-sample networks is the first single-sample approach that estimates each sample's complete network rather than simply re-purposing differential-expression information for network-based analysis. More importantly, LIONESS fills a critical gap, enabling the predictions made by existing network reconstruction methodologies to be evaluated using the same statistical techniques widely applied in other areas of genomic data analysis. The mathematical framework of LIONESS is highly generalizable and has the potential to be used to study many different and important questions in the fields of precision medicine, health and biomedical research.

\section{Competing interests}\label{competing}
None of the authors have any competing interests.

\section{Acknowledgments}\label{acknowledgments}
This work was supported by grants from the US National Heart Lung Blood Institute of the National Institutes of Health (R01HL111759, P01HL105339, K25HL133599) and from the Charles A. King Trust Postdoctoral Research Fellowship Program, Sara Elizabeth O'Brien Trust, Bank of America, N.A., Co-Trustees. We also would like to thank Farrah Roy, Abhijeet Sonawane, and John Platig for useful insights and suggestions in drafting this manuscript.

\section{Data availability}\label{data_availability}
The data sets analyzed during the current study are available from the Gene Expression Omnibus under accession number GSE4987 (yeast data), GSE19480 (human RNA-seq data), and GSE31388 (human DNase data). The human data is also available online at \url{http://eqtl.uchicago.edu/}. The {\it in silico} data we generated are available from the authors upon request. Fully processed and normalized versions of the yeast and human data used in this study are also available from the authors upon request.

\bibliographystyle{apsrev-title}
\bibliography{lionessbib} 

      \counterwithin{equation}{section}
      \setcounter{equation}{0}
      \renewcommand{\theequation}{S\arabic{equation}}

\clearpage
\section{Supplemental Materials and Methods}\label{onlinemethods}
This document contains additional information regarding the data processing and analyses presented in the main text of ``Estimating Sample-Specific Regulatory Networks''. A summary of the data and analyses described in this supplement, and presented in both the main text, figures, and supplemental figures, is presented in Supplemental Table~\ref{suptab2} at the end of this document.

\subsection{Current single-sample analysis approaches}\label{supmeth_current}
Several existing methods quantify the expression differences associated with a single-sample (as compared to a background set of samples) and use that information in a network-type of analysis. These approaches all use differential analyses---either differential-expression or differential-correlation---to highlight information specific to a single sample. In other words, by design, they cannot estimate network structures common across all samples. These common structures are critical in analyzing networks and are needed to correctly quantify topological characteristics such as community structure, as well as node and edge centralities~\ref{supref_sonawane2017}. An overview of existing single-sample analysis approaches is included below and the methods are summarized in Supplemental Table~\ref{suptab1} at the end of this document.

\subsubsection*{Single-sample differential-expression}
The main way others have used single-sample information in a network analysis, is to start with a single ``known'' network and then overlay sample-specific expression information to identify the parts of this network that may be relevant in a sample-specific context. For these approaches, a single-sample differential-expression (ssDE) profile is first constructed by comparing the expression of genes in a given sample with an expected distribution of expression values across a background set of samples. One common way to quantify ssDE is using a Z-score approach. In this case, the mean and standard deviation of a gene's expression across samples is calculated; the expression of that gene in a given sample is then normalized by subtracting the mean and dividing by the standard deviation. Single-sample differential-expression has been used in network-analysis approaches in two main ways.\\

{\it Single-sample gene-set enrichment}: Both ssMARINa (single-sample Master Regulator Inference Algorithm)~\ref{supref_aytes2014} and VIPER (Virtual Inference of Protein activity by Enriched Regulon)~\ref{supref_alvarez2016} use ``sample-specific signatures'' obtained from ssDE analysis together with profiles for transcription factor targets to estimate the overall activity of transcriptional regulators and/or proteins in individual samples. However, these methods do not provide an estimate of the actual single-sample networks that may be leading to this differential activity.

{\it ssDE layered onto an input network}: In DERA (Differentially Expressed Regulation Analysis)~\ref{supref_liu2015}, a prior biological network is built from public databases, the various sample-specific portions of this network are estimated by ``coloring'' genes based on the ssDE results. These sample-specific subnetworks are analyzed to identify a core set of interactions commonly-identified across a group of samples. We note that using this type of approach, an edge that is specific to an individual sample, but not present in this original prior network, will never be identified. One can imagine that these interactions may be biologically important, such as when a mutation causes a protein to change its interacting partners~\ref{supref_wang2015}.

\subsubsection*{Single-sample differential-correlation}
Another way others have used single-sample expression information in a network analysis is to apply a statistical approach to quantify single-sample differential-correlation (ssPCC)~\ref{supref_zhang2014},\ref{supref_liu2016}. Pearson correlation follows a normal distribution, therefore the difference between two distributions of Pearson correlations can be statistically quantified using the Z-score. In other words, by calculating the Pearson correlation both with and without a sample of interest, this known relationship can be used to transform the difference between those correlations into a value representing the ssPCC.

One mathematical assumption made by ssPCC is that every edge has the same distribution of weight-values across the predicted single-sample models. In other words, all edges have equal probability of being identified across the population. Therefore, to remove false positives and generate interpretable results, ssPCC has been used to differentially-weight edges in a known biological network (e.g. documented protein-protein interactions in StringDB)~\ref{supref_liu2016}. As with DERA (see above), we point out that by doing this filtering, an edge that is specific to an individual sample, but not present in this original prior network, will never be identified.

\subsubsection*{Contrast with the LIONESS approach}
We emphasize that none of the above approaches for analyzing single-sample information are designed to directly estimate sample-specific networks. Both ssDE are ssPCC are conceptually quite similar. Importantly, by only quantifying how specific a node (ssDE) or edge (ssPCC) is to a specific sample, both approaches effectively mask any network relationships that may be common across the samples. In contrast, LIONESS is designed to estimate both sample-specific and common network relationships. In other words, LIONESS is fundamentally different from these existing single-sample analysis methods in that it estimates each sample's complete network rather than simply re-purposing differential-expression information for network-based analysis.

Furthermore, we point out that the approaches outlined above ignore both the extensive literature on network reconstruction methods, as well as the fact that regulatory networks are often characterized by nonlinear dynamics and synergistic effects. This is especially true of ssPCC, which is simply a re-framing of the residuals obtained from running a linear Pearson correlation analysis.

\subsection{Aggregate network reconstruction approaches}\label{supmeth_networks}

Many methods have been developed for inferring biological networks. In our manuscript we analyze the application of LIONESS to four specific methods: (1) Pearson correlation, (2) PANDA (Passing Attributes between Networks for Data Assimilation), (3) mutual information, and (4) Context Likelihood of Relatedness (CLR). These methods were chosen because they represent a set of network reconstruction methods that use either a linear (Pearson) or nonlinear (mutual information) correlation measure, and methods that extend those measures to try to better capture regulatory interactions.

{\it Pearson Correlation}: Pearson correlation evaluates the degree of a linear relationship between two variables. Regulatory networks can be reconstructed by calculating the Pearson correlation coefficient between the expression levels of each TF and each target gene. These coefficients are measures of whether TFs and target genes are being co-expressed, which may indicate a regulatory event.

{\it Passing Attributes between Networks for Data Assimilation (PANDA)}: PANDA~\ref{supref_glass2013} builds regulatory networks by starting with a prior of possible interactions between TFs and target genes, for example TF motif binding information. PANDA integrates this regulatory prior with gene expression information and protein-protein interaction data, using a message passing approach to determine information flow between the different data types. The message passing algorithm used in PANDA is based on the assumption that if the expression of two genes correlate, those genes are more likely to be regulated by similar sets of TFs than two genes that do not show correlation in expression.

{\it Mutual Information (MI)}: MI compares the joint probability distribution of two variables to the products of their corresponding marginal distributions. Similar to Pearson correlation, MI is a measure of association between TFs and target genes, which may indicate a regulatory event. This method does not assume linearity or continuity of the data used for building the network.

{\it Context Likelihood of Relatedness (CLR)}: CLR~\ref{supref_faith2007} is based on MI, but applies a double Z-score transformation to the MI scores to produce a ``normalized'' value for each edge ($z_{ij}$). This transformation normalizes each TF-gene interaction based on the background distribution of MI values for each gene ($z_j$), and the background distribution of MI values for each TF ($z_i$): $z_{ij}=\sqrt{z_i^2+z_j^2}$; note that any edge for which either $z_i<0$ or $z_j<0$ is given a final weight of zero ($z_{ij}=0$).

\subsubsection*{Running LIONESS on networks reconstructed using PANDA}
We used the {\it corr()} function and a version of the PANDA algorithm implemented in MATLAB~\ref{supref_glass2015} to reconstruct networks using Pearson correlation and PANDA, respectively. PANDA requires a prior regulatory network in addition to gene expression information. For the {\it in silico} data we used an identity matrix (corresponding to each transcription factor targeting only itself) as the prior regulatory network. For the yeast and human data, a prior regulatory network was constructed based on transcription sequence-motif information (see below). It is worth noting that although PANDA can optionally take protein-protein interaction (PPI) information as an input, we did not use PPI data to provide a fair comparison with other network reconstruction approaches.

\subsubsection*{Running LIONESS on networks reconstructed using mutual information and CLR}
To calculate the mutual information and reconstruct networks based on the Context Likelihood of Relatedness (CLR), we used the {\it build.mim()} and {\it clr()} functions within the ``minet'' package in R/Bioconductor~\ref{supref_meyer2008}. For the mutual information application in {\it in silico} data we used an estimator based on the entropy of the empirical probability distribution (estimator=``mi.empirical''), with 100 bins of equal width (disc=``equalwidth''); the application in yeast used default parameters. It is worth noting that these algorithms create symmetric gene by gene matrices of predicted edge scores. Therefore, in the context of reconstructing single-sample yeast {\it regulatory} networks, we reduced these aggregate networks by selecting only the portion of the predicted gene-by-gene matrix that corresponds to edges from a transcription factor to a target gene.

\subsection{Building an intuition for LIONESS using Pearson correlation and mutual information}\label{supmeth_intuition}

One advantage of the LIONESS approach is that the input edge-weight estimates can, in theory, come from any network inference method that leverages information across a set of samples. With this in mind, in order to gain a better understanding and intuition for LIONESS, we performed a detailed exploration of its behavior when the aggregate network model is calculated using two widely-used measures: Pearson correlation and mutual information. In particular, to gain a better intuition for the values estimated by LIONESS for these two measures, we simulated data for pairs of nodes that had either (1) a strong linear ($\rho=0.9699$), or (2) a strong nonlinear ($MI=1.34595$) relationship across multiple samples and applied LIONESS to these data.

In the linear case (Supplemental Figure~\ref{supfig1}A), we used Pearson correlation as our aggregate model and applied the LIONESS equation (Equation~\ref{main_Eq4} in the main text) to estimate edge-weights for each sample in the data. We observe that all samples are given similar edge-weights by LIONESS, with an average value of 0.9702. We then repeated this analysis for (1) a smaller number of samples, (2) increased noise, and (3) the introduction of outliers that are inconsistent with the expected linear relationship. We find that the edge-weights estimated by LIONESS are very robust to sample-size. Increasing the noise does decrease the edge-weight estimates for the samples that are farthest from the linear trend-line, but this is a desired outcome as these samples are the least consistent with the expected relationship. Similarly, when we add in samples to the data that are, by design, inconsistent with the expected linear relationship, LIONESS correctly identifies and gives strong negative edge-weights to those outlier samples.

In the nonlinear case (Supplemental Figure~\ref{supfig1}B) we used mutual information to estimate the nodes' aggregate relationship before applying the LIONESS equation. We find that mutual information is a slightly more sensitive measure than Pearson correlation, with samples that differ from the expected relationship more readily identified and down-weighted by LIONESS. Other than this, however, the conclusions from the mutual information and Pearson correlation analyses are very similar. LIONESS is highly robust to the number of samples used in both cases, and gives higher edge-weights to samples near the expected relationship but lower weights to the samples that are inconsistent with the overall expected relationship.

\subsection{Data generation, normalization, and relevant pre-processing steps}\label{supmeth_data}

\subsubsection*{Generation of the {\rm in silico} expression data and regulatory networks}
{\it Data generation}: To test LIONESS's ability to reconstruct sample-specific network models we generated a set of network models and a corresponding associated set of gene expression profiles. To begin, we created a single ``seed'' network model with $M$ nodes. To approximate the structure of biological networks~\ref{supref_albert2005}, the out-degree of nodes in this seed model were given a power-law degree distribution (generated using the approach published in~\ref{supref_clauset2009}, with $\alpha=3$) with their targets selected randomly. We ensured that the out-degree and in-degree of all nodes in this seed network was greater than zero.

Next we randomized this ``seed'' network model, holding the degree distribution fixed, by performing $\rho\times N_e$ edge-swaps (where $N_e$ is the number of edges and $\rho$ allows us to control how different the permuted network is from the initial seed network). Then we generated a set of initial Boolean states for each node in the network, and determined the subsequent states of the nodes using Stouffer's Z-score method:

\begin{equation}\label{onlineEq1}
S_j^{(t+1)} = round \Bigg{[} CDF^{-1} \Bigg{(} \frac{\displaystyle\sum_i Z_{ij}S_{i}^{(t)}}{\sqrt{\displaystyle\sum_i S_{i}^{(t)}}} \Bigg{)} \Bigg{]}
\end{equation}

where $CDF^{-1}$ is the inverse cumulative distribution function for the normal distribution, $Z_{ij}$ is the Z-scored weight of an edge from node $i$ to node $j$, and $S_{i}^{(t)}$ is the state of node $i$ at time $t$. This Boolean model was run until an attractor solution was found. In total we generated 1000 random initial Boolean states for each randomized network, resulting in 1000 attractor solutions. The expression level of a node in the randomized network was then estimated as the average across these steady-state solutions. This entire process was then repeated $N$ times to create $N$ total randomized versions of the ``seed'' network model and $N$ corresponding matched expression samples. We applied this approach to generate sets of {\it in silico} networks that (1) are of various size (varying $M$), (2) have different levels of inter-network variability (varying $\rho$), (3) have varying levels of noise added to their associated expression data (see below), or (4) have a large number of samples (increasing $N$).

{\it Analysis with varying levels of edge permutation}: For our analyses, we created three initial ``seed'' networks of different sizes, with $M$ equal to 100, 250, and 625 nodes. To evaluate the impact of between-sample heterogeneity on LIONESS's prediction, for each of these seed networks we created six different sets of network-models based on different levels of edge-permutation, with $\rho$ equal to 0.5, 1, 1.5, 2, 2.5, or 3. Next, we ran the Boolean model described above on each of the generated networks. In total this process created 18 sets of $N=100$ ``gold-standard'' networks and the corresponding gene-expression levels for these 100 samples. These data sets represent networks of different sizes and permutation levels relative to the initial ``seed''. For each of these data sets, we constructed all 100 single-sample networks by applying LIONESS to aggregate networks based on the Pearson correlation in gene expression levels. We benchmarked these 100 single-sample networks against the 100 ``gold-standard'' networks in the data set. We also separately benchmarked only the ``permuted'' edges. To identify permuted edges we compared the edges in each of the gold-standards with the original ``seed'' network from which those standards were derived, and identified the subset of edges that only exist in either the ``seed'' or the ``gold-standard'' network. We then evaluated the AUC of the LIONESS-predicted single-sample edge-weights for this subset of edges. We observe that LIONESS estimates these edges incredibly well, with an overall accuracy similar to the other ``non-permuted'' edges. In contrast, the aggregate network is completely unable to estimate the ``permuted'' sample-specific edges, especially in cases of low heterogeneity (low values of $\rho$, which correspond to fewer edge-swaps).

{\it Analysis with varying levels of expression noise}: We took the data sets associated with an edge-permutation level of $\rho=1$. This included three data sets, representing network-sizes of $M$ equal to 100, 250, or 625 nodes. For each of these data sets, we determined the standard deviation ($\sigma_i$) of each gene's ($i$) expression across the samples. Then, for each gene, we used the {\it normrnd()} function in MATLAB to generate 100 noisy expression-values (one for each sample) based on a Gaussian distribution centered at that gene's original ``correct'' expression level in the sample, and with a standard deviation set to $r\times\sigma_i$. We did this for a range of values for $r$: 0.25, 0.5, 0.75, 1, 1.25, and 1.5. This resulted in eighteen additional {\it in silico} expression-sets (six levels of noise for the three data sets associated with different network sizes). Note that setting $r=0$ is equivalent to using the original expression-set without any additional noise. For each of these expression-sets we constructed the 100 corresponding single-sample networks by applying LIONESS to aggregate networks based on the Pearson correlation. We benchmarked both the full network models and the ``permuted edges'' in these 100 single-sample networks against the 100 ``gold-standard'' networks in the original data set. We find that as noise is added, the overall difference in AUC of the single-sample networks relative to the aggregate network model does decrease, but remains greater than zero. Importantly, we do not lose the ability to accurately predict the truly sample-specific ``permuted'' edges.

{\it Analysis with varying numbers of samples}: We selected the seed network associated with the $M=100$ nodes and created a data set with a moderate amount of heterogeneity among the networks ($\rho=1$) and containing $N=10,000$ samples. We selected subsets of these data consisting of various numbers of samples and applied LIONESS to aggregate networks build using Pearson correlation. We observe that including more samples increased the accuracy of the overall aggregate network, but that this corresponded to a poor prediction of sample-specific (permuted) edges. On the other hand, for the LIONESS single-sample networks, both the overall accuracy and the accuracy of the sample-specific edges is robust to the number of samples used. We also used this data set to assess the accuracy of LIONESS networks built using different aggregate reconstruction approaches: Pearson correlation, PANDA, mutual information, and CLR. For this final evaluation we applied each of the four reconstruction approaches to build four aggregate networks based on all 10,000 {\it in silico} expression samples. We then applied LIONESS to each of these aggregate networks, generating 10,000 sample-specific networks for each of the approaches.

\subsubsection*{Processing the yeast cell cycle expression data}
GPR files associated with~\ref{supref_pramila2006} were downloaded from the Gene Expression Omnibus (GEO; accession GSE4987). Each of two replicates were separately ma-normalized using the {\it maNorm()} function in the ``marray'' package in R/Bioconductor~\ref{supref_yang2009}. The data was batch-corrected using the {\it ComBat()} function in the ``sva'' package~\ref{supref_leek} and probe-sets mapping to the same gene were averaged, resulting in expression values for 5088 genes across fifty samples, twenty-five from each of the two replicate data sets. Two samples (corresponding to the 105 minute time point) were excluded for data-quality reasons, as noted in the original publication, and genes without motif information (see below) were then removed, giving a final expression data-set containing 48 samples and 3551 genes. These data were quantile-normalized and used in all subsequent analyses.

\subsubsection*{Generating the yeast motif prior data for PANDA}
PANDA requires a prior regulatory network structure in addition to gene expression information. To construct a motif prior network for yeast we downloaded predicted binding sites for 204 yeast transcription factors~\ref{supref_harbison2004}--\ref{supref_macisaac2006}. These data include 4360 genes with tandem promoters. 3551 of these genes are also covered on the gene expression array (see above). 105 total transcription factors in this data set target the promoter of one of these 3551 genes. The motif map between these 105 transcription factors and 3551 target genes was used as a prior regulatory network input to the PANDA algorithm. A subset of 65 of these transcription factors that also had expression information was used to reduce the size of the Pearson, MI, and CLR predicted networks to edges that extend between transcription factors and genes.

\subsubsection*{Processing the human RNA-Seq data}
RNA sequencing (RNA-Seq) data~\ref{supref_pickrell2010} were downloaded from the Pritchard lab website (\url{http://eqtl.uchicago.edu/}, accessed April 2014; also available on the Gene Expression Omnibus, GEO: GSE19480). 173 different samples corresponding to 74 different cell lines were available for download. We aligned all samples to the hg19/GRCh37 reference genome using Bowtie~\ref{supref_langmead2009}, with options -n 3 and -m 1, allowing for not more than 3 mismatches in the seed (28 bases on the high-quality end of the read, the default in Bowtie), and suppressing all non-unique alignments. We used RNA-SeQC~\ref{supref_deluca2012} to determine the quality of the reads, using an expression profiling efficiency cut-off of 0.75. Cell lines NA19119 and NA18853 fell below this cut-off. Next we used subread~\ref{supref_liao2013} to count reads, and the subread algorithm featureCounts to assign and summarize counts to genes. Finally, we removed samples with poor quality reads, and samples for which we did not have good quality DNase hypersensitivity data available (see below), leaving us with 153 samples corresponding to 65 different cell lines.

We used the ``DESeq2''~\ref{supref_love2014} package to analyze read counts for these 153 samples and adjusted for different library sizes using the {\it estimateSizeFactors} function. Only genes that had raw counts in at least 50\% of all 153 samples (21516/57820, or 37\% of all genes) were retained for further analysis. Correction for gene length was also performed; each intensity value was divided by the length of the corresponding gene (defined as the total length of the genomic region covered by the features/exons) in the ``gene\_id'' meta-feature in featureCounts. Finally, Ensembl gene ids were converted to HGNC gene symbols (16901/21516 genes) using R package ``biomaRt''~\ref{supref_durinck2005},\ref{supref_durinck2009}. This gene list was subsetted to only include genes for which we found at least one transcription factor binding site in the adjacent promoter (see below), and for which we had at least one DNase hypersensitivity peak-call in the adjacent promoter (see below). This resulted in a matrix of 12424 HGNC symbols by 153 expression samples.

\subsubsection*{Generating the human motif prior for PANDA}
We downloaded JASPAR motifs (\url{http://jaspar.genereg.net/})~\cite{mathelier2013jaspar} and then used Haystack~\ref{supref_pinello2014} to scan the entire hg19 genome for these motifs. Of the 205 motifs in the JASPAR database, only 158 had genomic hits that met our significance threshold ($p<10^{-6}$). We used HOMER~\ref{supref_heinz2010} (\url{http://homer.salk.edu/homer/ngs/index.html}) to get the distances of these motif hits to the nearest transcriptional start site (TSS). We used these reported distances to parse the motif hits based on their TSS proximity, keeping only those hits within the ``promoter'' region of a gene, which we define as $[-750,+250]$ around the TSS. We then filtered for genes to include only those in the RNA-seq data (see above). This information was used to make a prior transcription factor to gene map that we used when running PANDA.

\subsubsection*{Processing the human DNase hypersensitivity data}
Raw DNAse hypersensitivity data~\ref{supref_degner2012} was downloaded from the Pritchard lab website (\url{http://eqtl.uchicago.edu/dsQTL\_data/RAW\_DATA/}, accessed June 2014; also available on GEO: GSE31388). A total of 204 different samples corresponding to 70 different cell lines were available for download. Data were aligned to the hg19/GRCh37 reference genome using Bowtie~\ref{supref_langmead2009}, with options -v 1 and -m 10, allowing for not more than 1 mismatch, and suppressing any alignments for reads having more than 10 reportable alignments. Quality control using Bowtie output identified 16 samples with a high percentage (greater than 80\%) of failed reads. We called DNase hypersensitivity peaks using MACS~\ref{supref_zhang2008}. Peaks with significance score of less than $10^{-5}$ were mapped to the nearest gene using HOMER~\ref{supref_heinz2010}. When the peak fell within the promoter region of the gene, we assigned a score to that sample-gene pair equal to $-10\cdot log_{10}p$; otherwise the sample-gene pair was given a default score of zero. We then removed the 16 poor quality samples, as well as samples for which we did not have good quality RNA-seq data (see above), leaving us with 177 samples corresponding to 65 cell lines.

To obtain a promoter-DNase score specific for each cell line, we averaged technical replicates. We then filtered these data to include only genes for which we also had RNA-seq data available and genes that were present in our motif prior (see above). The result was a matrix of DNase-promoter scores that included 12424 genes that had a DNase promoter-peak in at least one sample (at least one row-entry greater than zero). 3488 genes had a promoter-peak in all samples (all row-entries greater than zero).

\subsection{Analysis of the human single-sample networks}\label{supmeth_analysis}

\subsubsection*{Calculating gene degree and comparing with DNase hypersensitivity data}
When comparing with the DNase information, we calculated gene degree three different ways:

\begin{align}
\label{onlineEq2}
 {k}_{j}^{(L)}&=\displaystyle\sum_{i} p_{ij}^{(L)};\\
 {k}_{j}^{(L+e)}&=\displaystyle\sum_{i} {p}_{ij}^{(L)}{p}_{i}^{(e)};\\
 {k}_{j}^{(m+e)}&=\displaystyle\sum_{i} {p}_{ij}^{(m)}{p}_{i}^{(e)}
\end{align}
%
where $p_{i}^{(e)}$ is the ``probability'' that TF $i$ is expressed, calculated by taking the inverse CDF of the Z-score of the TF's expression compared to the background of its expression in all other samples; $p_{ij}^{(L)}$ is the ``probability'' of the edge from the LIONESS networks, found by taking the inverse CDF of the predicted edge-weight score (which is in Z-score units); $p_{ij}^{(m)}$ is the ``probability'' of the edge from the motif data, either 0 or 1 based on whether the motif of TF $i$ was found in the promoter of gene $j$.

\subsubsection*{Clustering networks/expression and running GSEA}
We created clusters of networks and expression samples by clustering either edge-weights, or gene expression levels, respectively. To do this we performed a hierarchical clustering in which we row-normalized edge-weights (or gene expression) across samples, calculated distance based on the Spearman correlation, and performed a complete-linkage clustering. Spearman correlation was used as a distance metric because the network edges are often not normally distributed. For each clustering performed, we took the primary cut of the dendrogram to make exactly two groups of samples for further analysis.

We performed a LIMMA~\ref{supref_smyth2004} analysis to identify either differentially-expressed or differentially-targeted genes between the two groups of samples defined by the hierarchical clustering. A gene's targeting was defined as the sum of all edges pointing to that gene in that sample (the gene's ``in-degree''). We also calculated the log fold-change in gene expression/targeting between the two groups of samples and ran a pre-ranked GSEA analysis with 1000 iterations. For differences in gene-targeting in network-defined clusters we observed 68 significant Reactome pathway signatures~\ref{supref_liberzon2011} with an FDR $<0.1$ and a gene set size less than forty. An equivalent clustering/LIMMA/GSEA analysis evaluating differential-expression on expression-defined clusters resulted in no significant Reactome pathways.

\subsection{Supplemental References}\label{supmeth_refs}
\begin{small}
\begin{enumerate}[leftmargin=0.65cm, label={[\arabic*]}, ref={[\arabic*]}]
    \item\label{supref_sonawane2017}A. R. Sonawane, J. Platig, M. Fagny, C.-Y. Chen, J. N. Paulson, C. M. Lopes-Ramos, D. L. DeMeo, J. Quackenbush, K. Glass, and M. L. Kuijjer, ''Understanding tissue-specifc gene regulation,`` Cell reports {\bf 21}, 1077 (2017).
    \item\label{supref_aytes2014}A. Aytes, A. Mitrofanova, C. Lefebvre, M. J. Alvarez, M. Castillo-Martin, T. Zheng, J. A. Eastham, A. Gopalan, K. J. Pienta, M. M. Shen, et al., ''Crossspecies regulatory network analysis identifies a synergistic interaction between foxm1 and cenpf that drives prostate cancer malignancy,`` Cancer cell {\bf 25}, 638 (2014).
    \item\label{supref_alvarez2016}M. J. Alvarez, Y. Shen, F. M. Giorgi, A. Lachmann, B. B. Ding, B. H. Ye, and A. Califano, ''Functional characterization of somatic mutations in cancer using networkbased inference of protein activity,`` Nature genetics {\bf 48}, 838 (2016).
    \item\label{supref_liu2015}C. Liu, R. Louhimo, M. Laakso, R. Lehtonen, and S. Hautaniemi, ''Identifcation of sample-specific regulations using integrative network level analysis,`` BMC cancer {\bf 15}, 319 (2015).
    \item\label{supref_wang2015}Y. Wang, N. Sahni, and M. Vidal, ''Global edgetic rewiring in cancer networks,`` Cell systems {\bf 1}, 251 (2015).
    \item\label{supref_zhang2014}W. Zhang, T. Zeng, and L. Chen, ''Edgemarker: identifying differentially correlated molecule pairs as edgebiomarkers,`` Journal of theoretical biology {\bf 362}, 35 (2014).
    \item\label{supref_liu2016}X. Liu, Y.Wang, H. Ji, K. Aihara, and L. Chen, ''Personalized characterization of diseases using sample-specific networks,`` Nucleic acids research {\bf 44}, e164 (2016).
    \item\label{supref_glass2013}K. Glass, C. Huttenhower, J. Quackenbush, and G.-C.Yuan, ''Passing messages between biological networks to refine predicted interactions,`` PloS one {\bf 8}, e64832 (2013).
    \item\label{supref_faith2007}J. J. Faith, B. Hayete, J. T. Thaden, I. Mogno, J. Wierzbowski, G. Cottarel, S. Kasif, J. J. Collins, and T. S. Gardner, ''Large-scale mapping and validation of escherichia coli transcriptional regulation from a compendium of expression profiles,`` PLoS biology {\bf 5}, e8 (2007).
    \item\label{supref_glass2015}K. Glass, J. Quackenbush, and J. Kepner, ''High performance computing of gene regulatory networks using a message-passing model,`` High Performance Extreme Computing Conference (HPEC), 2015 IEEE pp. 1--6 (2015).
    \item\label{supref_meyer2008}P. E. Meyer, F. Lafitte, and G. Bontempi, ''minet: A R/bioconductor package for inferring large transcriptional networks using mutual information,`` BMC bioinformatics {\bf 9}, 461 (2008).
    \item\label{supref_albert2005}R. Albert, ''Scale-free networks in cell biology,`` Journal of cell science {\bf 118}, 4947 (2005).
    \item\label{supref_clauset2009}A. Clauset, C. R. Shalizi, and M. E. Newman, ''Powerlaw distributions in empirical data,`` SIAM review {\bf 51}, 661 (2009).
    \item\label{supref_pramila2006}T. Pramila, W. Wu, S. Miles, W. S. Noble, and L. L. Breeden, ''The forkhead transcription factor hcm1 regulates chromosome segregation genes and fills the s-phase gap in the transcriptional circuitry of the cell cycle,`` Genes \& development {\bf 20}, 2266 (2006).
    \item\label{supref_yang2009}Y. Yang, A. Paquet, and S. Dudoit, ''marray: Exploratory analysis for two-color spotted microarray data,`` Version 1.16 (2007).
    \item\label{supref_leek}J. T. Leek, W. E. Johnson, H. S. Parker, A. E. Jaffe, and J. D. Storey, Package 'sva'".
    \item\label{supref_harbison2004}C. T. Harbison, D. B. Gordon, T. I. Lee, N. J. Rinaldi, K. D. Macisaac, T. W. Danford, N. M. Hannett, J.-B. Tagne, D. B. Reynolds, J. Yoo, et al., ''Transcriptional regulatory code of a eukaryotic genome,`` Nature {\bf 431}, 99 12 (2004).
    \item\label{supref_fraenkel}Fraenkel Lab, ''Regulatory Map formatted for spreadsheet import,``, \url{http://fraenkel.mit.edu/Harbison/release_v24/txtfiles/} (2004).
    \item\label{supref_macisaac2006}K. D. MacIsaac, T. Wang, D. B. Gordon, D. K. Gifford, G. D. Stormo, and E. Fraenkel, ''An improved map of conserved regulatory sites for saccharomyces cerevisiae,`` BMC bioinformatics {\bf 7}, 113 (2006).
    \item\label{supref_pickrell2010}J. K. Pickrell, J. C. Marioni, A. A. Pai, J. F. Degner, B. E. Engelhardt, E. Nkadori, J.-B. Veyrieras, M. Stephens, Y. Gilad, and J. K. Pritchard, ''Understanding mechanisms underlying human gene expression variation with rna sequencing,`` Nature {\bf 464}, 768 (2010).
    \item\label{supref_langmead2009}B. Langmead, C. Trapnell, M. Pop, and S. L. Salzberg, ''Ultrafast and memory-efficient alignment of short dna sequences to the human genome,`` Genome Biol {\bf 10}, R25 (2009).
    \item\label{supref_deluca2012}D. S. DeLuca, J. Z. Levin, A. Sivachenko, T. Fennell, M.-D. Nazaire, C. Williams, M. Reich, W. Winckler, and G. Getz, ''Rna-seqc: Rna-seq metrics for quality control and process optimization,`` Bioinformatics {\bf 28}, 1530 (2012).
    \item\label{supref_liao2013}Y. Liao, G. K. Smyth, and W. Shi, ''The subread aligner: fast, accurate and scalable read mapping by seed-andvote,`` Nucleic acids research p. gkt214 (2013).
    \item\label{supref_love2014}M. I. Love, W. Huber, and S. Anders, ''Moderated estimation of fold change and dispersion for rna-seq data with deseq2,`` Genome biology {\bf 15}, 550 (2014).
    \item\label{supref_durinck2005}S. Durinck, Y. Moreau, A. Kasprzyk, S. Davis, B. De Moor, A. Brazma, and W. Huber, ''Biomart and bioconductor: a powerful link between biological databases and microarray data analysis,`` Bioinformatics {\bf 21}, 3439 (2005).
    \item\label{supref_durinck2009}S. Durinck, P. T. Spellman, E. Birney, and W. Huber, ''Mapping identifiers for the integration of genomic datasets with the r/bioconductor package biomart,`` Nature protocols {\bf 4}, 1184 (2009). 
    \item\label{supref_mathelier2014}A. Mathelier, X. Zhao, A. W. Zhang, F. Parcy, R. Worsley-Hunt, D. J. Arenillas, S. Buchman, C.-y. Chen, A. Chou, H. Ienasescu, et al., ''Jaspar 2014: an extensively expanded and updated open-access database of transcription factor binding profiles,`` Nucleic acids research p. gkt997 (2013).
    \item\label{supref_pinello2014}L. Pinello, J. Xu, S. H. Orkin, and G.-C. Yuan, ''Analysis of chromatin-state plasticity identifies cell-type-specific regulators of h3k27me3 patterns,`` Proceedings of the National Academy of Sciences {\bf 111}, E344 (2014).
    \item\label{supref_heinz2010}S. Heinz, C. Benner, N. Spann, E. Bertolino, Y. C. Lin, P. Laslo, J. X. Cheng, C. Murre, H. Singh, and C. K. Glass, ''Simple combinations of lineage-determining transcription factors prime cis-regulatory elements required for macrophage and b cell identities,`` Molecular cell {\bf 38}, 576 (2010).
    \item\label{supref_degner2012}J. F. Degner, A. A. Pai, R. Pique-Regi, J.-B. Veyrieras, D. J. Gaffney, J. K. Pickrell, S. De Leon, K. Michelini, N. Lewellen, G. E. Crawford, et al., ''Dnase I sensitivity qtls are a major determinant of human expression variation,`` Nature {\bf 482}, 390 (2012).
    \item\label{supref_zhang2008}Y. Zhang, T. Liu, C. A. Meyer, J. Eeckhoute, D. S. Johnson, B. E. Bernstein, C. Nusbaum, R. M. Myers, M. Brown, W. Li, et al., ''Model-based analysis of chipseq (macs),`` Genome Biol {\bf 9}, R137 (2008).
    \item\label{supref_smyth2004}G. K. Smyth, ''Linear models and empirical bayes methods for assessing differential expression in microarray experiments,`` Statistical applications in genetics and molecular biology {\bf 3}, 1 (2004).
    \item\label{supref_liberzon2011}A. Liberzon, A. Subramanian, R. Pinchback, H. Thorvaldsd D\'ottir, P. Tamayo, and J. P. Mesirov, ''Molecular signatures database (msigdb) 3.0,`` Bioinformatics {\bf 27}, 1739 (2011).
  \end{enumerate}
\end{small}

\clearpage

\onecolumngrid 
  \subsection{Summary of data and analyses}\label{supmeth_summary}

    \counterwithin{figure}{subsection}
    \setcounter{figure}{0}
    \renewcommand{\thefigure}{S\arabic{figure}}
    \renewcommand{\figurename}{Supplemental Table}

    \begin{figure*}[htbp]
      \includegraphics[width=500px]{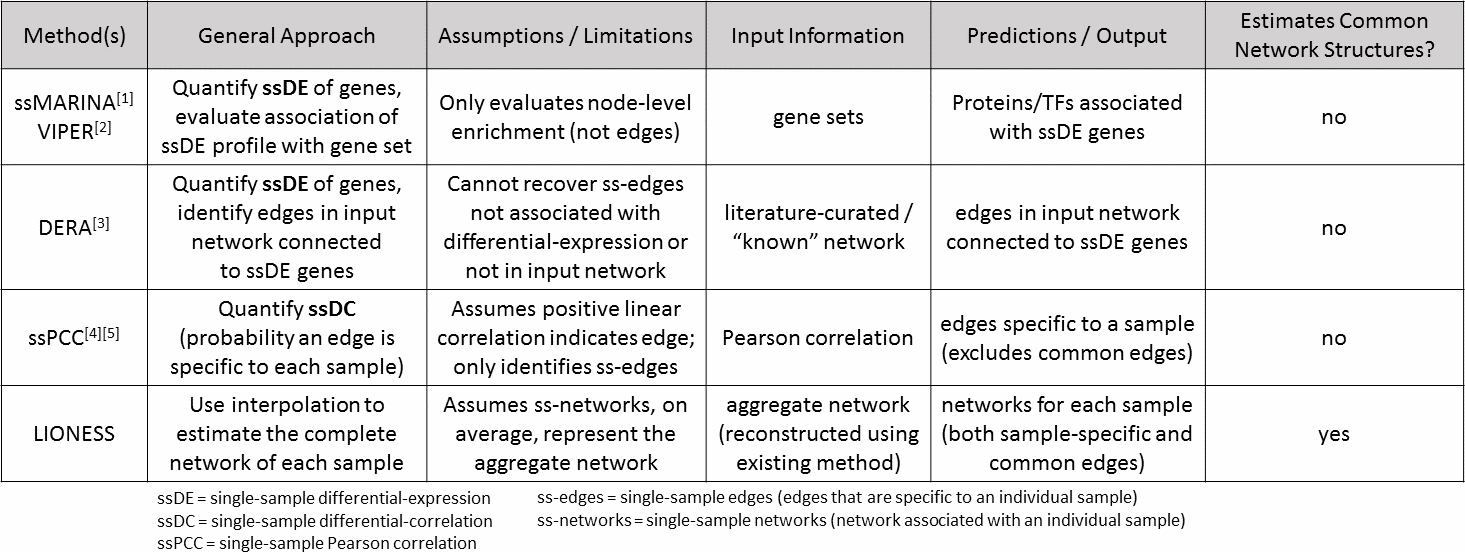}
      \caption{A summary of single-sample analysis approaches.}
      \label{suptab1}
    \end{figure*}

    \begin{figure*}[htbp]
      \includegraphics[width=500px]{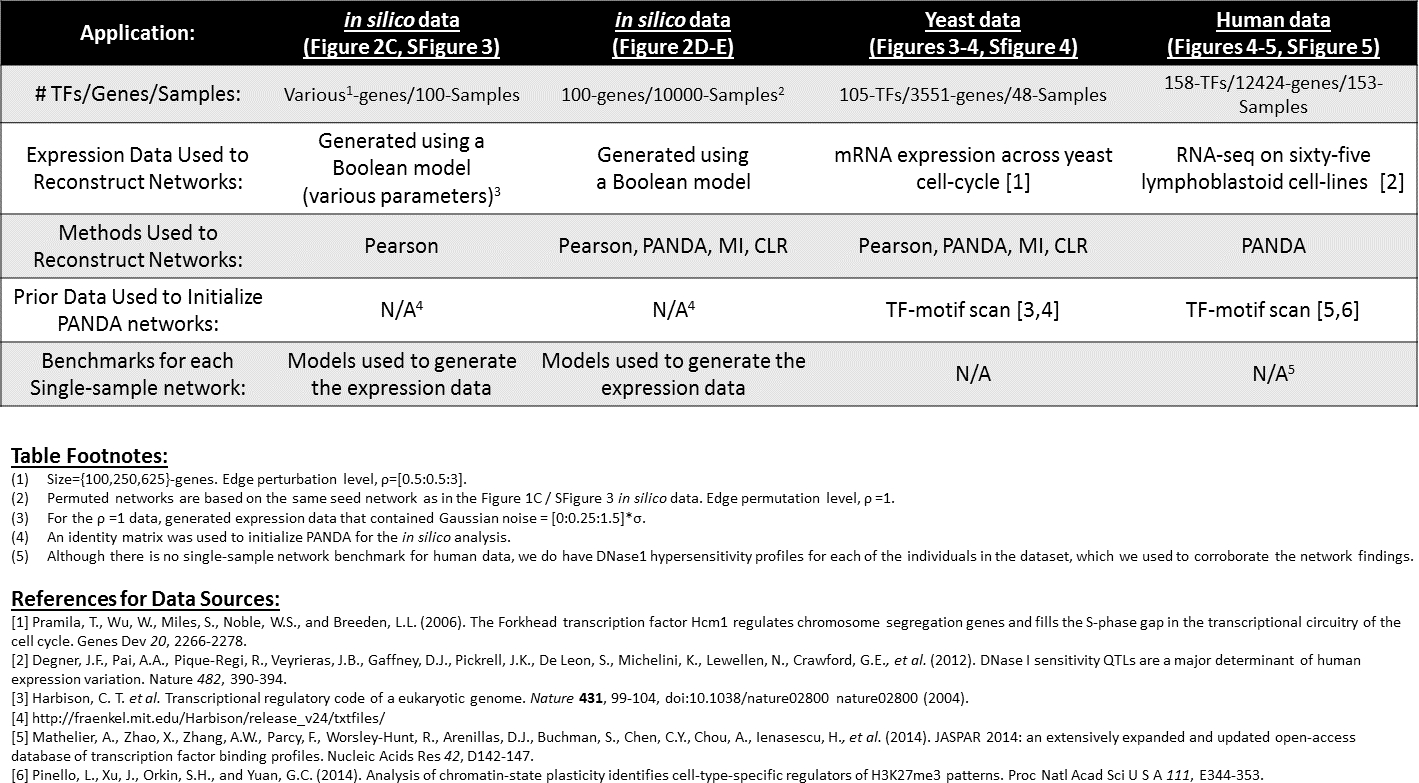}
      \caption{A summary of the analyses we have performed in this manuscript and the data used for each.}
      \label{suptab2}
    \end{figure*}

\clearpage
    \subsection{Supplemental Figures}\label{supmeth_figs}
    \begin{subappendices}

      \counterwithin{figure}{subsection}
      \setcounter{figure}{0} 
      \renewcommand{\thefigure}{S\arabic{figure}} 
      \renewcommand{\figurename}{Supplemental Figure}
    \begin{figure*}[htbp]
      \includegraphics[width=500px]{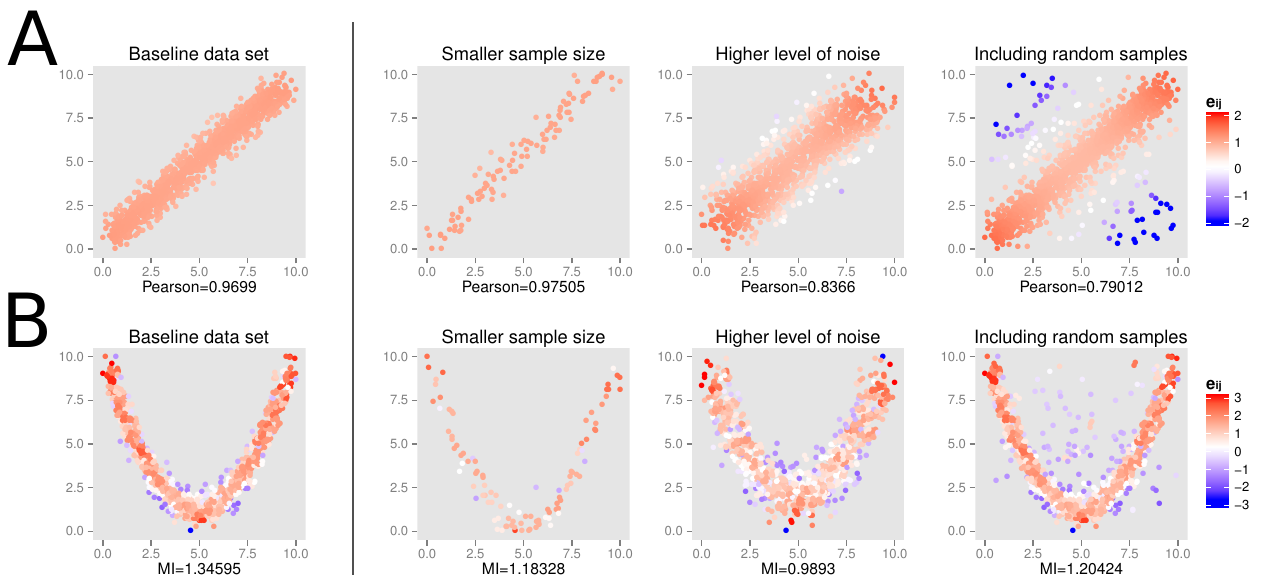}
      \caption{Examples of single-sample edge weights estimated by applying LIONESS to (A) Pearson correlation or (B) mutual information (MI). Plots show simulated values for two variables that represent the expression levels for two genes across a set of samples. For (A) the samples follow a linear relationship, $x=y$, and for (B) they follow a nonlinear relationship, $y=10\cdot(x/5-1)^2$. Each dot corresponds to an individual sample. A dot's color indicates the edge weight ($e_{ij}$) estimated for that sample using LIONESS (red: positive edge weights, blue: negative edge weights, see color legends). The aggregate statistics are shown below the plots. The plots on the left of the vertical bar show a ``baseline'' data set consisting of $N=1000$ samples with a low level of Gaussian noise (standard deviation, $sd=0.1$ for (A) and $sd=0.05$ for (B)) added to the $x$ versus $y$ relationship. The three plots to the right of the vertical bar show data sets with (1) a lower number of input samples ($N=100$), (2) a higher level of Guassian noise ($sd=0.25$ for (A), $sd=0.1$ for (B)), or (3) additional, randomly-distributed outlier samples ($0.25\cdot N$ for (A), $0.1\cdot N$ for (B)). Outliers were generated based on Gaussian noise with mean $=5$, $sd=3$, and bounded by $[0,10]$ using the ``truncnorm'' package in R.}
      \label{supfig1}
    \end{figure*}
    \begin{figure*}[!ht]
      \includegraphics[width=500px]{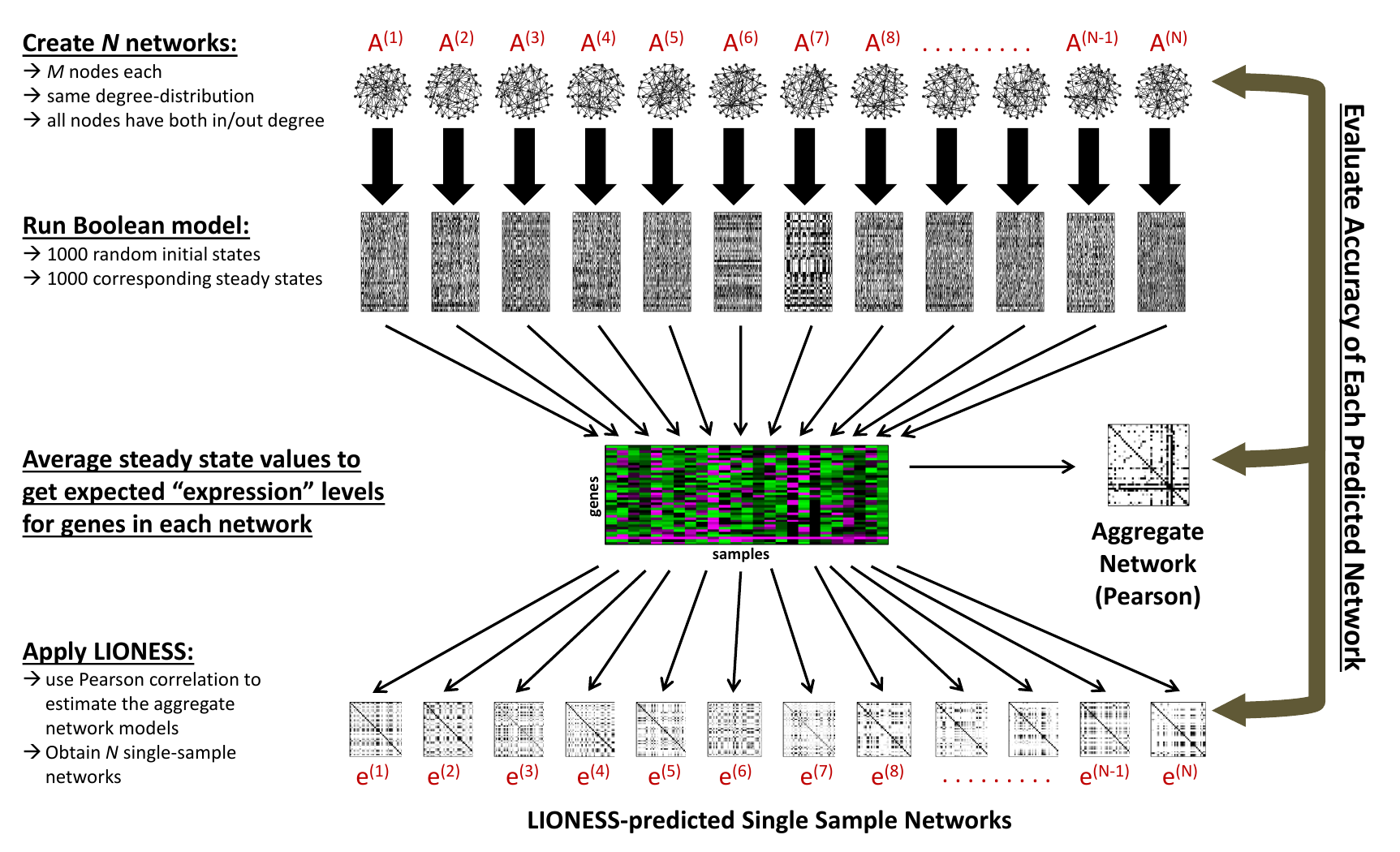}
      \caption{A schematic overview of how we generated {\it in silico} expression data for a set of known underlying gene regulatory network models.}
      \label{supfig2}
    \end{figure*}
    \begin{figure*}[!ht]
      \includegraphics[width=500px]{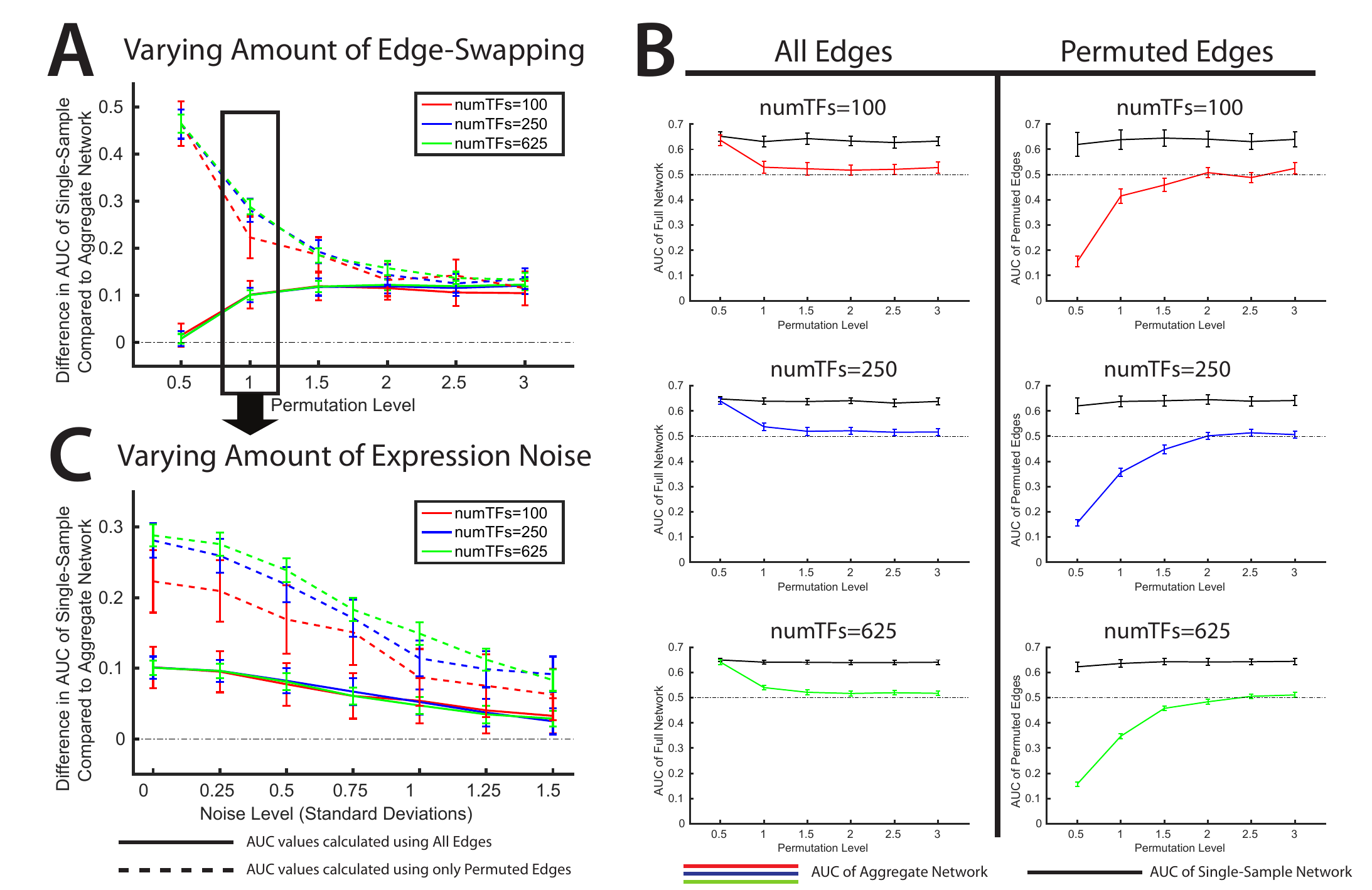}
      \caption{Results from applying LIONESS to in silico data sets. (A) A plot showing the difference in AUC values for the aggregate versus the LIONESS single-sample networks across different networks sizes and for different permutation levels. Solid lines show the evaluation across all edges in the network models, dashed lines show the evaluation when only considering edges that are ``permuted'' (different from the original ``seed'' network model). The mean and standard deviation across the 100 samples are shown. (B) Plots showing the average AUC values for the single-sample (black) and aggregate (colored) network models across different levels of edge-permutation. The range of AUC values, based on the standard deviation, is indicated by the error bars. The left panel shows the evaluation using all edges and the right shows the evaluation using only ``permuted'' edges. We see that the aggregate network models do a very poor job at accurately predicting the permuted edges, which are the edges that are truly sample-specific. (C) A plot showing the difference in AUC values between the aggregate and single-sample networks for different networks sizes and with different levels of Gaussian noise added to the expression information. Solid lines show the evaluation across all edges in the network models, dashed lines show the evaluation when only considering ``permuted'' edges. The mean and standard deviation across the 100 samples are shown.}
      \label{supfig3}
    \end{figure*}
    \begin{figure*}[!ht]
      \includegraphics[width=500px]{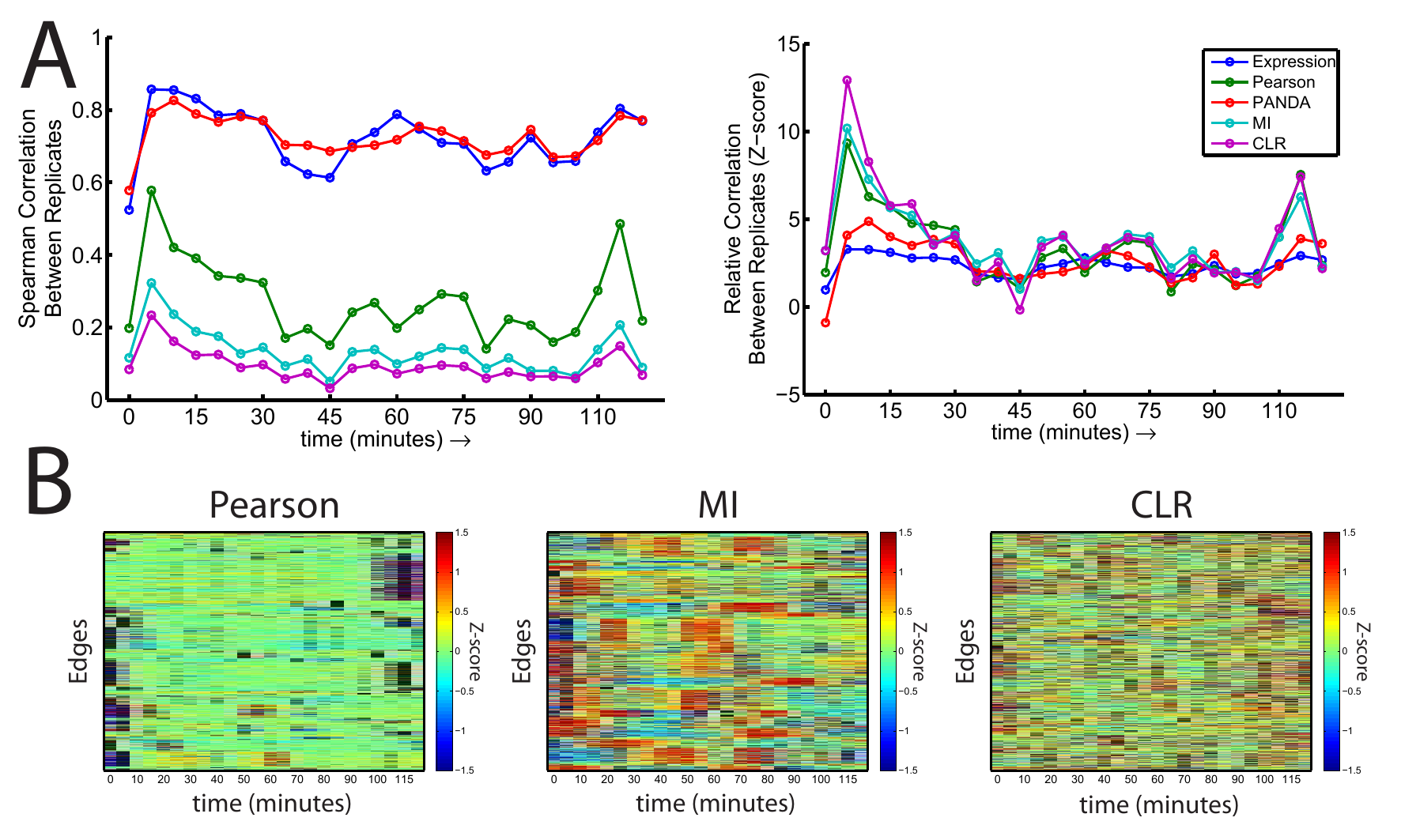}
      \caption{(A) The left panel shows the similarity, as determined by the Spearman correlation, between replicates in the expression data, and the similarity between networks corresponding to those same replicates in the ``R1-from-R1 \& R2-from-R2'' reconstruction. The right panel shows the relative similarity. In this case the Spearman correlation values from the left panel have been converted into Z-scores, based on the mean and standard deviation across the correlation values between all possible pairs of samples (compare with Figure~\ref{fig3} in the main text). (B) Heat maps showing the top 1000 most variables edges identified when applying LIONESS to yeast cell cycle data using various other aggregate reconstruction approaches. Although LIONESS is a generalizable approach, these results highlight the importance of selecting a robust underlying aggregate reconstruction algorithm when applying LIONESS.}
      \label{supfig4}
    \end{figure*}
    \begin{figure*}[!ht]
      \includegraphics[width=500px]{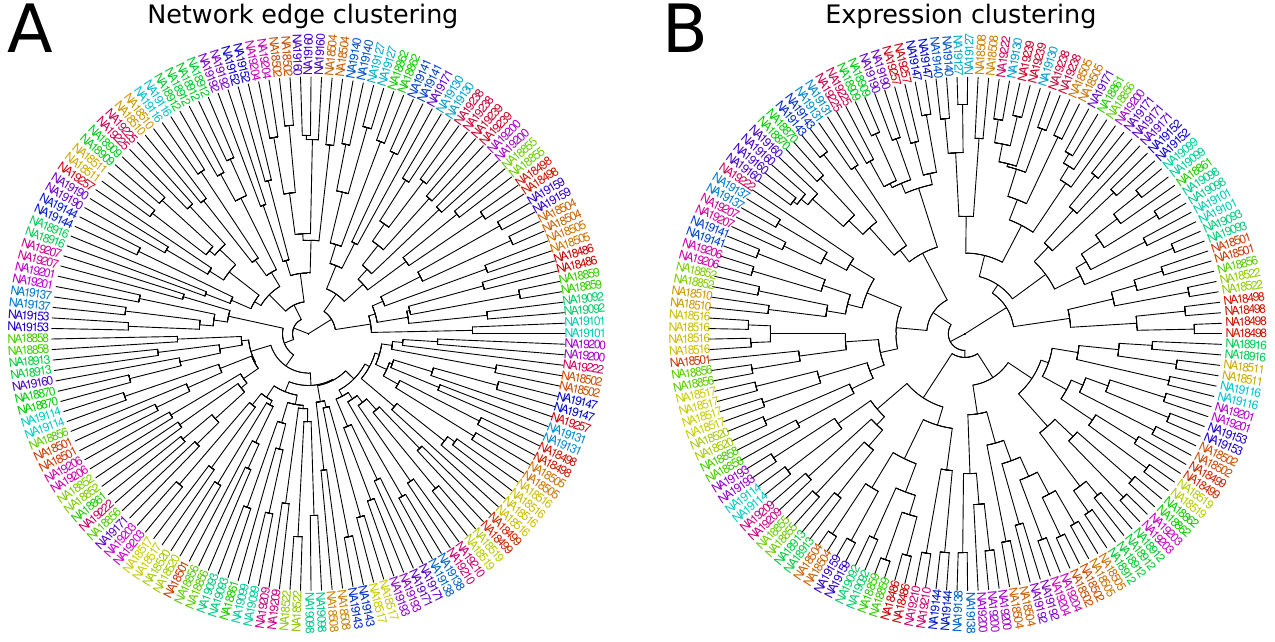}
      \caption{A hierarchical clustering of (A) the 153 single-sample lymphoblastoid cell line networks predicted using LIONESS and (B) the gene expression data that was used to build the networks. Closer inspection reveals that technical replicates (different experimental samples assaying the same cell line) tend to cluster together. Different replicates cluster together in the network and expression dendrograms; this includes 121/153 of the single-sample networks and 121/153 of gene expression samples.}
      \label{supfig5}
    \end{figure*}
    \end{subappendices}

\clearpage
\begin{landscape}
\section{LIONESS: Linear Interpolation to Obtain Network Estimates for Single Samples}\label{SuppMat}

  \pagestyle{empty}
  \newcommand{\Lpagenumber}{\ifdim\textwidth=\linewidth\else\bgroup
    \dimendef\margin=0 
    \ifodd\value{page}\margin=\oddsidemargin
    \else\margin=\evensidemargin
    \fi
    \raisebox{\dimexpr -\topmargin-\headheight-\headsep-0.5\linewidth}[0pt][0pt]{%
      \rlap{\hspace{\dimexpr \margin+\textheight+\footskip}%
      \llap{\rotatebox{90}{\thepage}}}}%
  \egroup\fi}
  \AddEverypageHook{\Lpagenumber}%

\newcommand\tab[1][1cm]{\hspace*{#1}}

  \counterwithin{equation}{section}
  \setcounter{equation}{0} 
  \renewcommand{\theequation}{E\arabic{equation}}

In section~\ref{section_derivation}, we show how we can derive networks for individual samples from an aggregate network by applying the LIONESS equation. The LIONESS equation works independently of the network reconstruction algorithm that is used to infer the aggregate network, and can be applied to both linear and non-linear network reconstruction algorithms. To demonstrate this, we show that we can exactly solve the LIONESS equation in its application to Pearson correlation networks, a linear network reconstruction algorithm, in section~\ref{section_pearson}. In addition, we show the applicability of the LIONESS equation to the non-linear Mutual Information network reconstruction algorithm in section~\ref{section_mi}. Finally, in~\ref{section_convergence}, we show that LIONESS networks do not converge in the limit of a large number of samples.

\subsection{Derivation to find regulatory networks for individual samples in a collection}\label{section_derivation}

To begin, we assume that the value of a given edge (${e}_{ij}^{({\alpha})}$) from a transcription factor ($i$) to a gene ($j$) predicted by a network reconstruction algorithm using a collection of samples ($\alpha$) is the linear combination of the value of that edge across networks specific to each of the input samples ($e_{ij}^{(s)}$), where ${w}_{s}^{({\alpha})}$ represents the relative contribution of sample ($s$):
\begin{equation}
{e}_{ij}^{({\alpha})}=\displaystyle\sum_{s=1}^{N} w_s^{(\alpha)} e_{ij}^{(s)}, \text{  where  } \displaystyle\sum_{s=1}^{N} w_s^{(\alpha)} = 1
\label{Eq1}
\end{equation}
Given this assumption, we can calculate two ``aggregate'' networks, one using all samples (${e}_{ij}^{(\alpha)}$), as described above, and the other using all but one of the samples (${e}_{ij}^{(\alpha-q)}$):
\begin{equation}
{e}_{ij}^{(\alpha-q)}=\displaystyle\sum_{s \neq q}^{N} w^{(\alpha-q)}_s e_{ij}^{(s)}, \text{  where  } \displaystyle\sum_{s \neq q}^{N} w^{(\alpha-q)}_s = 1
\label{Eq2}
\end{equation}
Now, subtracting these two ``aggregate'' network estimates we get:
\begin{align}
{e}_{ij}^{(\alpha)}-{e}_{ij}^{(\alpha-q)} &= \displaystyle\sum_{s=1}^{N} w_s^{(\alpha)} e_{ij}^{(s)} - \displaystyle\sum_{s \neq q}^{N} w^{(\alpha-q)}_s e_{ij}^{(s)} \\
 &= w^{(\alpha)}_q e_{ij}^{(q)} + \displaystyle\sum_{s \neq q}^{N} w_s^{(\alpha)} e_{ij}^{(s)} - \displaystyle\sum_{s \neq q}^{N} w^{(\alpha-q)}_s e_{ij}^{(s)} \\
 &= w^{(\alpha)}_q e_{ij}^{(q)} + \displaystyle\sum_{s \neq q}^{N} (w^{(\alpha)}_s-w^{(\alpha-q)}_s) e_{ij}^{(s)}
 \label{Eq3L}
 \end{align}
We can then solve for the network specific to a single sample, $e_{ij}^{(q)}$:
\begin{align}
e_{ij}^{(q)} &= \frac{1}{w^{(\alpha)}_q} \Bigg{[} {e}_{ij}^{(\alpha)} - {e}_{ij}^{(\alpha-q)} + \displaystyle\sum_{s \neq q}^{N} (w^{(\alpha-q)}_s-w^{(\alpha)}_s) e_{ij}^{(s)} \Bigg{]} \\
&= \frac{1}{w^{(\alpha)}_q} \Bigg{[} {e}_{ij}^{(\alpha)} - \displaystyle\sum_{s \neq q}^{N} w^{(\alpha)}_s e_{ij}^{(s)} \Bigg{]}
\label{Eq4L}
\end{align}
If we then stipulate that the weights used to estimate ${e}_{ij}^{(\alpha)}$ are related to the weights used to estimate ${e}_{ij}^{(\alpha-q)}$ by a constant, $w^{(\alpha)}_s = C w^{(\alpha-q)}_s$ and combine with Equations \ref{Eq1} and \ref{Eq2}, we observe that:
\begin{equation}
1 = \displaystyle\sum_{s \neq q}^{N} w^{(\alpha-q)}_s = \displaystyle\sum_{s=1}^{N} w_s^{(\alpha)} = w_q^{(\alpha)}+ \displaystyle\sum_{s \neq q}^{N} w^{(\alpha)}_s
= w_q^{(\alpha)}+ C \displaystyle\sum_{s \neq q}^{N} w^{(\alpha-q)}_s = w_q^{(\alpha)} + C
\label{EqForC}
\end{equation}
This then implies that $C=1-w^{(\alpha)}_q$ (or, equivalently, $w^{(\alpha)}_q=1-w^{(\alpha)}_s / w^{(\alpha-q)}_s$), which we can substitute into Equation \ref{Eq4L} in order to calculate the values of the edges in the single-sample network, $e_{ij}^{(q)}$, in terms of the two ``aggregate'' networks:
\begin{align}
e_{ij}^{(q)} &= \frac{1}{w^{(\alpha)}_q} \Bigg{[} {e}_{ij}^{(\alpha)} - C \displaystyle\sum_{s \neq q}^{N} w^{(\alpha-q)}_s e_{ij}^{(s)} \Bigg{]} \\
 &= \frac{1}{w^{(\alpha)}_q} \Bigg{[} {e}_{ij}^{(\alpha)} - (1-w^{(\alpha)}_q) {e}_{ij}^{(\alpha-q)} \Bigg{]} \\
 &= \frac{1}{w^{(\alpha)}_q} \Bigg{[} {e}_{ij}^{(\alpha)} - {e}_{ij}^{(\alpha-q)} \Bigg{] + {e}_{ij}^{(\alpha-q)} }
\label{Eq5}
\end{align}
Finally, we can simplify this equation by (optionally) assuming that each sample is given equal weight ($w^{(\alpha)}_q=\frac{1}{N}$):
\begin{equation}
e_{ij}^{(q)} = N \Bigg{[} {e}_{ij}^{(\alpha)} - {e}_{ij}^{(\alpha-q)} \Bigg{] + {e}_{ij}^{(\alpha-q)} }
\label{LIONESS}
\end{equation}

\break
\subsection{Application to Pearson correlation}\label{section_pearson}
To begin, we remember that the Pearson correlation ($r$) between two variables, $X$ and $Y$ can be defined as:
\begin{equation}
r=\frac{1}{N-1}\displaystyle\sum_i^{N} \Bigg{(} \frac{X_i-\bar{X}}{S_X} \Bigg{)} \Bigg{(} \frac{Y_i-\bar{Y}}{S_Y}\Bigg{)},
\text{  where  } \bar{X}=\frac{1}{N}\displaystyle\sum_i^{N} X_i
\text{  and  } S_X=\sqrt{\frac{1}{N-1}\displaystyle\sum_i^{N} (X_i-\bar{X})^2}
\label{Pearson}
\end{equation}
We can then use the Pearson correlation to calculate two ``aggregate'' networks, one using all samples (resulting in $r$), and the other using all samples except for sample $q$ (resulting in $r'$):
\begin{equation}
r'=\frac{1}{N-2}\displaystyle\sum_{i\neq q}^{N} \Bigg{(} \frac{X_i-\bar{X'}}{S'_X} \Bigg{)} \Bigg{(} \frac{Y_i-\bar{Y'}}{S'_Y}\Bigg{)},
\text{  where  } \bar{X'}=\frac{1}{N-1}\displaystyle\sum_{i \neq  q}^{N} X_i
\text{  and  } S'_X=\sqrt{\frac{1}{N-2}\displaystyle\sum_{i \neq q}^{N} (X_i-\bar{X'})^2}
\label{PearsonPrime}
\end{equation}
Now, using the ``LIONESS'' equation for deriving single samples we see that:
\begin{align}
e_{ij}^{(q)} &= N (e_{ij}^{(\alpha)}-e_{ij}^{(\alpha - q)})+e_{ij}^{(\alpha - q)} & & \\
r_{xy}^{(q)} &= N (r_{xy}-r'_{xy})+r'_{xy} & & \\
&= N \Bigg{[}\frac{1}{N-1}\displaystyle\sum_i^{N} \Bigg{(} \frac{X_i-\bar{X}}{S_X} \Bigg{)} \Bigg{(} \frac{Y_i-\bar{Y}}{S_Y}\Bigg{)}
- \frac{1}{N-2}\displaystyle\sum_{i\neq q}^{N} \Bigg{(} \frac{X_i-\bar{X'}}{S'_X} \Bigg{)} \Bigg{(} \frac{Y_i-\bar{Y'}}{S'_Y}\Bigg{)} \Bigg{]}
 + \frac{1}{N-2}\displaystyle\sum_{i\neq q}^{N} \Bigg{(} \frac{X_i-\bar{X'}}{S'_X} \Bigg{)} \Bigg{(} \frac{Y_i-\bar{Y'}}{S'_Y}\Bigg{)} \\
&= \frac{N}{N-1}\displaystyle\sum_i^{N} \Bigg{(} \frac{X_i-\bar{X}}{S_X} \Bigg{)} \Bigg{(} \frac{Y_i-\bar{Y}}{S_Y}\Bigg{)}
- \frac{N-1}{N-2}\displaystyle\sum_{i\neq q}^{N} \Bigg{(} \frac{X_i-\bar{X'}}{S'_X} \Bigg{)} \Bigg{(} \frac{Y_i-\bar{Y'}}{S'_Y}\Bigg{)}\\
&= \frac{N}{N-1}\Bigg{(} \frac{X_q-\bar{X}}{S_X} \Bigg{)} \Bigg{(} \frac{Y_q-\bar{Y}}{S_Y}\Bigg{)} +
\frac{N}{N-1}\displaystyle\sum_{i\neq q}^{N} \Bigg{(} \frac{X_i-\bar{X}}{S_X} \Bigg{)} \Bigg{(} \frac{Y_i-\bar{Y}}{S_Y}\Bigg{)}
- \frac{N-1}{N-2}\displaystyle\sum_{i\neq q}^{N} \Bigg{(} \frac{X_i-\bar{X'}}{S'_X} \Bigg{)} \Bigg{(} \frac{Y_i-\bar{Y'}}{S'_Y}\Bigg{)}\\
&= \frac{N}{N-1}\Bigg{(} \frac{X_q-\bar{X}}{S_X} \Bigg{)} \Bigg{(} \frac{Y_q-\bar{Y}}{S_Y}\Bigg{)} +
\displaystyle\sum_{i\neq q}^{N} \Bigg{[}\Bigg{(}\frac{N}{N-1}\Bigg{)}\Bigg{(} \frac{X_i-\bar{X}}{S_X} \Bigg{)} \Bigg{(} \frac{Y_i-\bar{Y}}{S_Y}\Bigg{)}
- \Bigg{(}\frac{N-1}{N-2}\Bigg{)} \Bigg{(} \frac{X_i-\bar{X'}}{S'_X} \Bigg{)} \Bigg{(} \frac{Y_i-\bar{Y'}}{S'_Y}\Bigg{)}\Bigg{]}
\label{Eq3}
\end{align}
Next, if we average over all possible values of $q$ we find that:
\begin{align}
\frac{1}{N} \displaystyle\sum_i^{N} r_{xy}^{(i)} &= \frac{1}{N-1} \displaystyle\sum_i^{N} \Bigg{(} \frac{X_i-\bar{X}}{S_X} \Bigg{)} \Bigg{(} \frac{Y_i-\bar{Y}}{S_Y}\Bigg{)} +
 \displaystyle\sum_i^{N} \displaystyle\sum_{i\neq q}^{N} \Bigg{[}\Bigg{(}\frac{1}{N-1}\Bigg{)}\Bigg{(} \frac{X_i-\bar{X}}{S_X} \Bigg{)} \Bigg{(} \frac{Y_i-\bar{Y}}{S_Y}\Bigg{)}
- \Bigg{(}\frac{N-1}{N(N-2)}\Bigg{)} \Bigg{(} \frac{X_i-\bar{X'}}{S'_X} \Bigg{)} \Bigg{(} \frac{Y_i-\bar{Y'}}{S'_Y}\Bigg{)}\Bigg{]}
\label{AveragePearson}
\end{align}
We also note that (from equations \ref{Pearson} and \ref{PearsonPrime} above):
\begin{align}
\bar{X} &=\frac{1}{N} \displaystyle\sum_i^{N} X_i \\
&= \frac{1}{N} X_q + \frac{1}{N} \displaystyle\sum_{i \neq q}^{N} X_i \\
&= \frac{1}{N} X_q + \frac{N-1}{N} \bar{X'} \\
&= \bar{X'}+\frac{1}{N}(X_q-\bar{X'})
\label{Eq4}
\end{align}
Thus when the difference between $X_q$ and $\bar{X'}$ is much less than the total number of samples being considered ($N$), $\bar{X'} \rightarrow \bar{X} $ and $S'_X \rightarrow S_X$. This is most likely for large values of $N$. In this limit Equation \ref{Eq3} can be simplified as follows:
\begin{align}
\lim_{N\to\infty} r_{xy}^{(q)} &= \lim_{N\to\infty} \Bigg{[} \frac{N}{N-1}\Bigg{(} \frac{X_q-\bar{X}}{S_X} \Bigg{)} \Bigg{(} \frac{Y_q-\bar{Y}}{S_Y}\Bigg{)} +
\displaystyle\sum_{i\neq q}^{N} \Bigg{[}\Bigg{(}\frac{N}{N-1}\Bigg{)}\Bigg{(} \frac{X_i-\bar{X}}{S_X} \Bigg{)} \Bigg{(} \frac{Y_i-\bar{Y}}{S_Y}\Bigg{)}
- \Bigg{(}\frac{N-1}{N-2}\Bigg{)} \Bigg{(} \frac{X_i-\bar{X'}}{S'_X} \Bigg{)} \Bigg{(} \frac{Y_i-\bar{Y'}}{S'_Y}\Bigg{)}\Bigg{]} \Bigg{]} \\
&= \Bigg{(} 1 \Bigg{)} \Bigg{(} \frac{X_q-\bar{X}}{S_X} \Bigg{)} \Bigg{(} \frac{Y_q-\bar{Y}}{S_Y}\Bigg{)} +
 \displaystyle\sum_{i\neq q}^{N}  \Bigg{[} \Bigg{(} 1 \Bigg{)} \Bigg{(} \frac{X_i-\bar{X}}{S_X} \Bigg{)} \Bigg{(} \frac{Y_i-\bar{Y}}{S_Y}\Bigg{)} -
\Bigg{(} 1 \Bigg{)} \Bigg{(} \frac{X_i-\bar{X}}{S_X} \Bigg{)} \Bigg{(} \frac{Y_i-\bar{Y}}{S_Y}\Bigg{)} \Bigg{]}\\
&= \Bigg{(} \frac{X_q-\bar{X}}{S_X} \Bigg{)} \Bigg{(} \frac{Y_q-\bar{Y}}{S_Y}\Bigg{)}
\label{Eq6}
\end{align}
Similarly Equation \ref{AveragePearson} can be simplified as follows:
\begin{align}
\lim_{N\to\infty} \frac{1}{N} \displaystyle\sum_i^{N} r_{xy}^{(i)}
&= \lim_{N\to\infty} \Bigg{[} \frac{1}{N-1} \displaystyle\sum_i^{N} \Bigg{(} \frac{X_i-\bar{X}}{S_X} \Bigg{)} \Bigg{(} \frac{Y_i-\bar{Y}}{S_Y}\Bigg{)} +
 \displaystyle\sum_i^{N} \displaystyle\sum_{i\neq q}^{N} \Bigg{[}\Bigg{(}\frac{1}{N-1}\Bigg{)}\Bigg{(} \frac{X_i-\bar{X}}{S_X} \Bigg{)} \Bigg{(} \frac{Y_i-\bar{Y}}{S_Y}\Bigg{)}
- \Bigg{(}\frac{N-1}{N(N-2)}\Bigg{)} \Bigg{(} \frac{X_i-\bar{X'}}{S'_X} \Bigg{)} \Bigg{(} \frac{Y_i-\bar{Y'}}{S'_Y}\Bigg{)}\Bigg{]} \Bigg{]} \\
&= \lim_{N\to\infty} \Bigg{[} \frac{1}{N-1} \displaystyle\sum_i^{N} \Bigg{(} \frac{X_i-\bar{X}}{S_X} \Bigg{)} \Bigg{(} \frac{Y_i-\bar{Y}}{S_Y}\Bigg{)} +
\Bigg{(}\frac{1}{N-1}\Bigg{)} \displaystyle\sum_i^{N} \displaystyle\sum_{i\neq q}^{N} \Bigg{[}\Bigg{(} \frac{X_i-\bar{X}}{S_X} \Bigg{)} \Bigg{(} \frac{Y_i-\bar{Y}}{S_Y}\Bigg{)}
- \Bigg{(}\frac{(N-1)^2}{N(N-2)}\Bigg{)} \Bigg{(} \frac{X_i-\bar{X'}}{S'_X} \Bigg{)} \Bigg{(} \frac{Y_i-\bar{Y'}}{S'_Y}\Bigg{)}\Bigg{]} \Bigg{]} \\
&= \frac{1}{N-1} \displaystyle\sum_i^{N} \Bigg{(} \frac{X_i-\bar{X}}{S_X} \Bigg{)} \Bigg{(} \frac{Y_i-\bar{Y}}{S_Y}\Bigg{)} +
\Bigg{(} 0 \Bigg{)} \displaystyle\sum_i^{N} \displaystyle\sum_{i\neq q}^{N} \Bigg{[}\Bigg{(} \frac{X_i-\bar{X}}{S_X} \Bigg{)} \Bigg{(} \frac{Y_i-\bar{Y}}{S_Y}\Bigg{)}
- \Bigg{(} 1 \Bigg{)} \Bigg{(} \frac{X_i-\bar{X}}{S_X} \Bigg{)} \Bigg{(} \frac{Y_i-\bar{Y}}{S_Y}\Bigg{)}\Bigg{]} \\
&= \frac{1}{N-1} \displaystyle\sum_i^{N} \Bigg{(} \frac{X_i-\bar{X}}{S_X} \Bigg{)} \Bigg{(} \frac{Y_i-\bar{Y}}{S_Y}\Bigg{)} +
\Bigg{(} 0 \Bigg{)} \displaystyle\sum_i^{N} \displaystyle\sum_{i\neq q}^{N} \Bigg{(} 0 \Bigg{)} \\
&= \frac{1}{N-1}  \displaystyle\sum_i^{N}  \Bigg{(} \frac{X_q-\bar{X}}{S_X} \Bigg{)} \Bigg{(} \frac{Y_q-\bar{Y}}{S_Y}\Bigg{)}
\label{Eq7}
\end{align}
Which is, reassuringly, equal to $r$ (equation \ref{Pearson}). We note that in order to accurately obtain this final equation we must use the fact that, although $1/(N-1) \rightarrow 0$ for $N \rightarrow \infty$, here $1/(N-1)$ is multiplied by a summation over $N$ finite values. Thus, when we take the limit of large $N$ we have a very large number divided by another very large number of the same order of magnitude, resulting in a finite value.

\break
\subsection{Application to mutual information}\label{section_mi}
Mutual information is a way to capture non-linear relationships between two variables ($a$ and $b$) and can be used to define a network relationship $a \rightarrow b$ when there is a strong dependence (high mutual information) between those variables. In order to estimate the network for a single-sample, the LIONESS equation makes a linear assumption concerning the relationship of an edge $a \rightarrow b$ in one network (represented by the quantity $e_{ab}^{(1)}$), with the edge $a \rightarrow b$ in {\it another} network (represented by the quantity $=e_{ab}^{(2)}$). In other words, it only cares about the relationship between the quantities $e_{ab}^{(1)}$ and $e_{ab}^{(2)}$); it doesn't stipulate anything about the linear (or non-linear) relationship between $a$ and $b$ within either of those networks or the method that was used to derive the quantities $e_{ab}^{(1)}$ and $e_{ab}^{(2)}$.

Here we show that the linear assumption of LIONESS makes about the relationship of an edge between two networks holds true in the limit of large number of samples (large $N$), even when the magnitude of that edge $a \rightarrow b$ is calculated using a non-linear measure -- in this case mutual information. We do this by explicitly calculating the form of the mutual information for a single-sample based on the LIONESS equation. We then explicitly test the linear assumption made by LIONESS, and show that averaging over all such single-sample networks reduces to the aggregate mutual-information network in the limit of a large number of total samples (large $N$). The derivation presented in this section follows the same basic approach as the one testing the application of LIONESS to networks derived using Pearson correlation (see section~\ref{section_pearson}).

To begin, we remember that for $\mathcal{X}$ discrete states captured by variable $x$, and $\mathcal{Y}$ discrete states captured by variable $y$, mutual information can be defined as:
\begin{equation}
I(x,y)=\displaystyle\sum_{x \in \mathcal{X}} \displaystyle\sum_{y \in \mathcal{Y}} p(x,y) log \bigg{(} \frac{p(x,y)}{p(x) p(y)} \bigg{)}
\label{MIorg}
\end{equation}

In practice, in order to estimate $p(x)$, $p(y)$ and $p(x,y)$ samples will be binned into the $\mathcal{X}$ and $\mathcal{Y}$ discrete states. To better understand how this works, let us assume we are trying to find the mutual information between two vectors of data: $\vec{a}$ and $\vec{b}$. Based on this we can define functions $x_i=f_x(\vec{a}_i)$ and $y_i=f_y(\vec{b}_i)$ to return the binned ``state'' of each sample (or data-point) $i$ in the $(x,y)$ bins in $\mathcal{X}$ and $\mathcal{Y}$ space.

Using these functions we can then define a matrix $A_{xy}$ whose entries are defined as total number of data-points binned into each $(x,y)$ state:
\begin{equation}
A_{xy}=\displaystyle\sum_i \delta(x_i,x)\delta(y_i,y)
\label{Axy}
\end{equation}
where $\delta()$ is the Kronecker delta function. Let us also define:
\begin{equation}
X_{y}=\displaystyle\sum_i \delta(y_i,y) \text{   and   } Y_{x}=\displaystyle\sum_i \delta(x_i,x).
\end{equation}
Note that $X_y$ and $Y_x$ are the row and column sums of $A_{xy}$, respectively, and represent the total number of data points binned into each of the $(x)$ and $(y)$ states.

Based on the above definitions, we can re-write the mutual information based on $A_{xy}$, $X_y$ and $Y_x$. More specifically, we note that $p(x,y)=\sfrac{A_{xy}}{N}$ , $p(x)=\sfrac{Y_x}{N}$ and $p(y)=\sfrac{X_y}{N}$. Subbing these into Equation \ref{MIorg}:
\begin{equation}
I(x,y)=\frac{1}{N} \displaystyle\sum_{x \in \mathcal{X}} \displaystyle\sum_{y \in \mathcal{Y}} A_{xy} log \bigg{(} \frac{A_{xy} N}{X_y Y_x} \bigg{)}
\label{MIxy}
\end{equation}
This form of mutual information sums over binned-states ($x$ and $y$), but one can also equivalently write a form that sums over data-points. For this we define $A_{xy}^{(i)}$ as value of the $A_{xy}$ matrix that contains data-point $i$:
\begin{equation}
A_{xy}^{(i)}=\displaystyle\sum_j \delta(x_j,x_i)\delta(y_j,y_i)
\label{Axyi}
\end{equation}
and similarly $Y_{x}^{(i)}=\displaystyle\sum_j \delta(x_j,x_i)$ and $X_{y}^{(i)}=\displaystyle\sum_j \delta(y_j,y_i)$ as the values of $Y_x$ and $X_y$ that include the data-point $i$.

\begin{figure}[ht]
\begin{centering}
\includegraphics[width=500px]{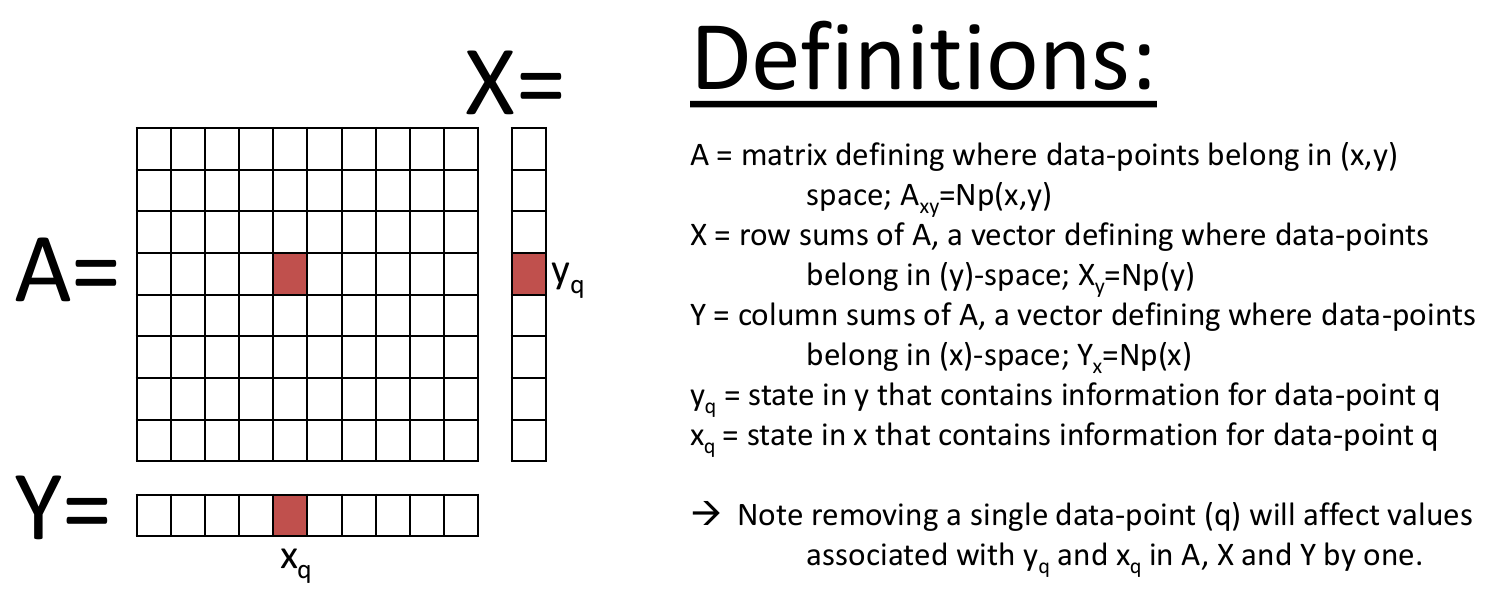}
\end{centering}
\end{figure}

We also note that in order to calculate mutual information as a sum over data points (instead of bin-states) the contribution of each data-point to the sum should be equivalent to one over the total number of data-points in the same $(x,y)$ bin-state as that data-point, or $A_{xy}^{(i)}$. Based on this information, we can re-write mutual information as:
\begin{equation}
I(x,y) =\frac{1}{N} \displaystyle\sum_{i} \frac{A_{xy}^{(i)}}{A_{xy}^{(i)}} log \bigg{(} \frac{A_{xy}^{(i)} N}{X_y^{(i)} Y_x^{(i)}} \bigg{)}
=\frac{1}{N} \displaystyle\sum_{i} log \bigg{(} \frac{A_{xy}^{(i)} N}{X_y^{(i)} Y_x^{(i)}} \bigg{)}
\label{MIi}
\end{equation}

Next, we define three new additional quantities. First, a matrix summarizing how many data-points would be assigned to each $(x,y)$ bin were we to remove a single data-point ($q$) from $\vec{a}$ and $\vec{b}$:
\begin{equation}
{A'}_{xy}^{(i)}=\displaystyle\sum_{j \neq q} \delta(x_j,x_i)\delta(y_j,y_i)=
\begin{cases}
A_{xy}^{(i)}-1 & x_i = x_q , y_i = y_q \\
A_{xy}^{(i)} & otherwise
\end{cases}
\label{Aprime}
\end{equation}
A vector summarizing how many data-points would be assigned to each $y$ state, were we to remove data-point $q$:
\begin{equation}
{X'}_{y}^{(i)}=\displaystyle\sum_{j \neq q} \delta(y_j,y_i)=
\begin{cases}
X_{y}^{(i)}-1 & y_i = y_q \\
X_{y}^{(i)} & otherwise
\end{cases}
\label{Xprime}
\end{equation}
And a vector summarizing how many data-points would be assigned to each $x$ state, were we to remove  data-point $q$:
\begin{equation}
{Y'}_{x}^{(i)}=\displaystyle\sum_{j \neq q} \delta(x_j,x_i)=
\begin{cases}
Y_{x}^{(i)}-1 & x_i = x_q \\
Y_{x}^{(i)} & otherwise
\end{cases}
\label{Yprime}
\end{equation}
The only stipulation in defining these quantities is that the same $x$ and $y$ states that were used to define $A_{xy}$, $X_y$ and $Y_x$ are also used to define ${A'}_{xy}$ , ${X'}_y$  and ${Y'}_x$.
Based on these quantities, we can then define the mutual information between two vectors $\vec{a}$ and $\vec{b}$ after removing a single data-point from those vectors ($q$, previously been assigned to bin $(x_q,y_q)$).
\begin{equation}
I'(x,y) =\frac{1}{N-1} \displaystyle\sum_{i \neq q} log \bigg{(} \frac{{A'}_{xy}^{(i)} N}{{X'}_y^{(i)} {Y'}_x^{(i)}} \bigg{)}
\end{equation}

Now, reiterating the LIONESS equation we see:
\begin{align}
I(q) &=N(I(x,y)-I'(x,y))+I'(x,y) \\
&=NI(x,y)-(N-1)I'(x,y) \\
&=\displaystyle\sum_{i} log \bigg{(} \frac{A_{xy}^{(i)} N}{X_y^{(i)} Y_x^{(i)}} \bigg{)} - \displaystyle\sum_{i \neq q} log \bigg{(} \frac{{A'}_{xy}^{(i)} N}{{X'}_y^{(i)} {Y'}_x^{(i)}} \bigg{)} \\
&=log \bigg{(} \frac{A_{xy}^{(q)} N}{X_y^{(q)} Y_x^{(q)}} \bigg{)} + \displaystyle\sum_{i \neq q} log \bigg{(} \frac{A_{xy}^{(i)} N}{X_y^{(i)} Y_x^{(i)}} \bigg{)} - \displaystyle\sum_{i \neq q} log \bigg{(} \frac{{A'}_{xy}^{(i)} N}{{X'}_y^{(i)} {Y'}_x^{(i)}} \bigg{)} \\
&=log \bigg{(} \frac{A_{xy}^{(q)} N}{X_y^{(q)} Y_x^{(q)}} \bigg{)} + \displaystyle\sum_{i \neq q} log \bigg{(} \frac{A_{xy}^{(i)} N}{X_y^{(i)} Y_x^{(i)}} \bigg{)} - log \bigg{(} \frac{{A'}_{xy}^{(i)} N}{{X'}_y^{(i)} {Y'}_x^{(i)}} \bigg{)} \\
&=log \bigg{(} \frac{A_{xy}^{(q)} N}{X_y^{(q)} Y_x^{(q)}} \bigg{)} + \displaystyle\sum_{i \neq q} log \bigg{(} \frac{N}{N-1} \frac{A_{xy}^{(i)}}{{A'}_{xy}^{(i)}}\frac{{X'}_y^{(i)}}{{X}_y^{(i)}} \frac{{Y'}_x^{(i)}}{{Y}_x^{(i)}}\bigg{)}
\label{LIONESSMI}
\end{align}
We note that, from above, we know the relationship between $A_{xy}^{(i)}$ and ${A'}_{xy}^{(i)}$ (see Equation \ref{Aprime}), the relationship between $X_y^{(i)}$ and ${X'}_y^{(i)}$(see Equation \ref{Xprime}), as well as the relationship between $Y_x^{(i)}$ and ${Y'}_x^{(i)}$ (see Equation \ref{Yprime}). Based on this information, let us divide the data-points in the right-hand sum of Equation \ref{LIONESSMI} into four parts: (1) $P_1$: those data-points that are not in either the $x_q$ or $y_q$ bins, (2) $P_2$: those data-points in the $x_q$ bin but not the $y_q$ bin, (3) $P_3$: those data-points in the $y_q$ bin but not the $x_q$ bin, and (4) $P_4$: those data-points that are with $q$ in the $(x_q,y_q)$ bin:
\begin{equation}
I(q) = log \bigg{(} \frac{A_{xy}^{(q)} N}{X_y^{(q)} Y_x^{(q)}} \bigg{)} + P_1 + P_2 + P_3 + P_4 .
\end{equation}

Let's start with the simplest scenario: the data-points that are not in either the $x_q$ or the $y_q$ bin ($P_1$). We can calculate the total number of data-points in this category as:
\begin{equation}
N_1=N-X_y^{(q)}-Y_x^{(q)}+A_{xy}^{(q)}
\end{equation}
Within the category the following also holds for these data-points: $A_{xy}^{(i)}={A'}_{xy}^{(i)}$, $X_y^{(i)}={X'}_y^{(i)}$, and $Y_x^{(i)}={Y'}_{x}^{(i)}$. This is true for all data-points $i$ that are not in either the $x_q$ or $y_q$ bins. Based on this, the $P_1$ reduces to:
\begin{equation}
P_1 = N_1 log \bigg{(}\frac{N}{N-1}\bigg{)}
\end{equation}

The second scenario ($P_2$) is a data-point that is in the $x_q$ bin, but not the $y_q$ bin. We can calculate the number of data-points in this category as:
\begin{equation}
N_2 = Y_x^{(q)}-A_{xy}^{(q)}
\end{equation}

\begin{figure}[ht]
\begin{centering}
\includegraphics[width=500px]{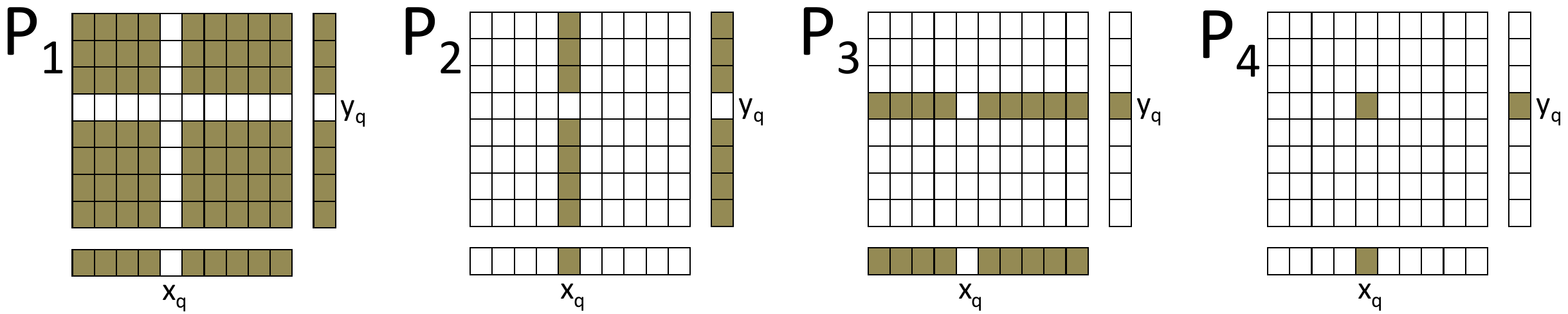}
\end{centering}
\end{figure}

Within the category the following also holds for these data-points: $A_{xy}^{(i)}={A'}_{xy}^{(i)}$, $X_y^{(i)}={X'}_y^{(i)}$, and ${Y'}_{x}^{(i)}={Y'}_{x}^{(q)}=Y_x^{(q)}-1$. In this case, the right hand sum in Equation \ref{LIONESSMI} reduces to:
\begin{equation}
P_2 = N_2 log \bigg{(}\frac{N}{N-1} \frac{Y_x^{(q)}-1}{Y_x^{(q)}}\bigg{)}
= N_2 \bigg{[}log \bigg{(}\frac{N}{N-1}\bigg{)}+log \bigg{(}\frac{Y_x^{(q)}-1}{Y_x^{(q)}}\bigg{)}\bigg{]}
\end{equation}

Similarly, the third scenario ($P_3$) is a data-point that is in the $y_q$ bin, but not in the $x_q$ bin. We can calculate the number of data-points in this category as:
\begin{equation}
N_3 = X_y^{(q)}-A_{xy}^{(q)}
\end{equation}
Within the category the following also holds for these data-points: $A_{xy}^{(i)}={A'}_{xy}^{(i)}$, ${X'}_y^{(i)}={X'}_y^{(q)}=X_y^{(q)}-1$, and $Y_x^{(i)}={Y'}_{x}^{(i)}$. In this case, the right hand sum in Equation \ref{LIONESSMI} reduces to:
\begin{equation}
P_3 = N_3 log \bigg{(}\frac{N}{N-1} \frac{X_y^{(q)}-1}{X_y^{(q)}}\bigg{)}
= N_3 \bigg{[}log \bigg{(}\frac{N}{N-1}\bigg{)}+log \bigg{(}\frac{X_y^{(q)}-1}{X_y^{(q)}}\bigg{)}\bigg{]}
\end{equation}

Finally, the last scenario is a data-point that is in the same $(x_q, y_q)$ bin as $q$ (but is not $q$ itself). The number of data-points meeting this criteria is precisely $A_{xy}^{(q)}-1$. The following also holds for these data-points: ${A'}_{xy}^{(i)}={A'}_{xy}^{(q)}={A}_{xy}^{(q)}-1$, ${X'}_y^{(i)}={X'}_y^{(q)}=X_y^{(q)}-1$, and ${Y'}_{x}^{(i)}={Y'}_{x}^{(q)}=Y_x^{(q)}-1$.
In this case, the right hand sum in Equation \ref{LIONESSMI} reduces to:
\begin{align}
P_4 &= (A_{xy}^{(q)}-1) log \bigg{(} \frac{N}{N-1} \frac{A_{xy}^{(q)}}{{A}_{xy}^{(q)}-1}\frac{{X}_y^{(q)}-1}{{X}_y^{(q)}} \frac{{Y}_x^{(q)}-1}{{Y}_x^{(q)}}\bigg{)} \\
&=(A_{xy}^{(q)}-1)\bigg{[}log \bigg{(}\frac{N}{N-1}\bigg{)}
+ log \bigg{(} \frac{A_{xy}^{(q)}}{{A}_{xy}^{(q)}-1}\bigg{)}
+ log \bigg{(}\frac{{X}_y^{(q)}-1}{{X}_y^{(q)}}\bigg{)}
+ log \bigg{(}\frac{{Y}_x^{(q)}-1}{{Y}_x^{(q)}}\bigg{)}
\bigg{]}
\end{align}
Now let's aggregate together all the components of the right-hand sum:
\begin{align}
P_1+P_2+P_3+P_4 &= (N-X_y^{(q)}-Y_x^{(q)}+A_{xy}^{(q)}) log \bigg{(}\frac{N}{N-1}\bigg{)}\\
&+ (Y_x^{(q)}-A_{xy}^{(q)}) \bigg{[}log \bigg{(}\frac{N}{N-1}\bigg{)}+log \bigg{(}\frac{Y_x^{(q)}-1}{Y_x^{(q)}}\bigg{)}\bigg{]}\\
&+ (X_y^{(q)}-A_{xy}^{(q)}) \bigg{[}log \bigg{(}\frac{N}{N-1}\bigg{)}+log \bigg{(}\frac{X_y^{(q)}-1}{X_y^{(q)}}\bigg{)}\bigg{]} \\
&+ (A_{xy}^{(q)}-1)\bigg{[}log \bigg{(}\frac{N}{N-1}\bigg{)}
+ log \bigg{(} \frac{A_{xy}^{(q)}}{{A}_{xy}^{(q)}-1}\bigg{)}
+ log \bigg{(}\frac{{X}_y^{(q)}-1}{{X}_y^{(q)}}\bigg{)}
+ log \bigg{(}\frac{{Y}_x^{(q)}-1}{{Y}_x^{(q)}}\bigg{)}\bigg{]}
\end{align}
Re-grouping we then find that:
\begin{align}
P_1+P_2+P_3+P_4 &=
(N-X_y^{(q)}-Y_x^{(q)}+A_{xy}^{(q)}+Y_x^{(q)}-A_{xy}^{(q)}+X_y^{(q)}-A_{xy}^{(q)}+A_{xy}^{(q)}-1)log \bigg{(}\frac{N}{N-1}\bigg{)} \\
&+(Y_x^{(q)}-A_{xy}^{(q)}+A_{xy}^{(q)}-1)log \bigg{(}\frac{Y_x^{(q)}-1}{Y_x^{(q)}}\bigg{)}\\
&+(X_y^{(q)}-A_{xy}^{(q)}+A_{xy}^{(q)}-1)log \bigg{(}\frac{X_y^{(q)}-1}{X_y^{(q)}}\bigg{)}\\
&+(A_{xy}^{(q)}-1)log \bigg{(} \frac{A_{xy}^{(q)}}{{A}_{xy}^{(q)}-1}\bigg{)}
\end{align}
Or simplified:
\begin{align}
P_1+P_2+P_3+P_4 &=
(N-1)log \bigg{(}\frac{N}{N-1}\bigg{)}
+(Y_x^{(q)}-1)log \bigg{(}\frac{Y_x^{(q)}-1}{Y_x^{(q)}}\bigg{)}\\
&+(X_y^{(q)}-1)log \bigg{(}\frac{X_y^{(q)}-1}{X_y^{(q)}}\bigg{)}
+(A_{xy}^{(q)}-1)log \bigg{(} \frac{A_{xy}^{(q)}}{{A}_{xy}^{(q)}-1}\bigg{)}
\end{align}
So then, remembering Equation \ref{LIONESSMI} we observe that:
\begin{align}
I(q)=log \bigg{(} \frac{A_{xy}^{(q)} N}{X_y^{(q)} Y_x^{(q)}} \bigg{)}
& +(N-1)log \bigg{(}\frac{N}{N-1}\bigg{)}
+(Y_x^{(q)}-1)log \bigg{(}\frac{Y_x^{(q)}-1}{Y_x^{(q)}}\bigg{)} \\
& +(X_y^{(q)}-1)log \bigg{(}\frac{X_y^{(q)}-1}{X_y^{(q)}}\bigg{)}
+(A_{xy}^{(q)}-1)log \bigg{(} \frac{A_{xy}^{(q)}}{{A}_{xy}^{(q)}-1}\bigg{)}
\end{align}
If we then average over all possible values of $q$, we find:
\begin{align}
\frac{1}{N} \displaystyle\sum_i^{N} I(i) = \frac{1}{N} \displaystyle\sum_i^{N} \bigg{[} log \bigg{(} \frac{A_{xy}^{(i)} N}{X_y^{(i)} Y_x^{(i)}} \bigg{)}
& +(N-1)log \bigg{(}\frac{N}{N-1}\bigg{)}
+(Y_x^{(q)}-1) log \bigg{(}\frac{Y_x^{(q)}-1}{Y_x^{(q)}}\bigg{)} \\
& +(X_y^{(q)}-1) log \bigg{(}\frac{X_y^{(q)}-1}{X_y^{(q)}}\bigg{)}
+(A_{xy}^{(q)}-1) log \bigg{(} \frac{A_{xy}^{(q)}}{{A}_{xy}^{(q)}-1}\bigg{)}\bigg{]}
\end{align}
We are most interested in the behavior of this quantity in the limit of a large number of samples ($N\rightarrow\infty$). In this case:
\begin{align}
\lim_{N\to\infty} \frac{1}{N} \displaystyle\sum_i^{N} I(i) =
\lim_{N\to\infty} \frac{1}{N} \displaystyle\sum_i^{N} \bigg{[} log \bigg{(} \frac{A_{xy}^{(i)} N}{X_y^{(i)} Y_x^{(i)}} \bigg{)}
& +(N-1)log \bigg{(}\frac{N}{N-1}\bigg{)}
+(Y_x^{(q)}-1) log \bigg{(}\frac{Y_x^{(q)}-1}{Y_x^{(q)}}\bigg{)} \\
& +(X_y^{(q)}-1) log \bigg{(}\frac{X_y^{(q)}-1}{X_y^{(q)}}\bigg{)}
+(A_{xy}^{(q)}-1) log \bigg{(} \frac{A_{xy}^{(q)}}{{A}_{xy}^{(q)}-1}\bigg{)}\bigg{]}
\end{align}
We note that the latter terms in the sum are constants (not dependent on $i$) and thus can be brought out of the sum. In addition, based on the rules of limits, we can divide the above equation into five separate limits:
\begin{align}
\lim_{N\to\infty} \frac{1}{N} \displaystyle\sum_i^{N} I(i) =
\lim_{N\to\infty} \frac{1}{N} \displaystyle\sum_i^{N} log \bigg{(} \frac{A_{xy}^{(i)} N}{X_y^{(i)} Y_x^{(i)}} \bigg{)}
& + \lim_{N\to\infty} (N-1)log \bigg{(}\frac{N}{N-1}\bigg{)}
+ \lim_{N\to\infty} (Y_x^{(q)}-1) log \bigg{(}\frac{Y_x^{(q)}-1}{Y_x^{(q)}}\bigg{)} \\
& + \lim_{N\to\infty} (X_y^{(q)}-1) log \bigg{(}\frac{X_y^{(q)}-1}{X_y^{(q)}}\bigg{)}
+\lim_{N\to\infty} (A_{xy}^{(q)}-1) log \bigg{(} \frac{A_{xy}^{(q)}}{{A}_{xy}^{(q)}-1}\bigg{)}
\label{limIq}
\end{align}
Solving each of these limits we first note that, based on Equation \ref{MIi}:
\begin{equation}
\lim_{N\to\infty} \frac{1}{N} \displaystyle\sum_i^{N} log \bigg{(} \frac{A_{xy}^{(i)} N}{X_y^{(i)} Y_x^{(i)}} \bigg{)} = \lim_{N\to\infty} I(x,y) = I(x,y)
\end{equation}
Next we observe that the remaining limits take two main forms. The limit applied to the $N$ and the $A_{xy}^{(q)}$ portions of the equation are of the form:
\begin{equation}
\lim_{z\to c} (z-1)log \bigg{(}\frac{z}{z-1}\bigg{)} = \lim_{z \to c} \frac{log(z)-log(z-1)}{(z-1)^{-1}}
\end{equation}
Then, applying l'Hopital's rule (and re-arranging, and applying l'Hopital again, etc):
\begin{align}
\lim_{z\to c} \frac{log(z)-log(z-1)}{(z-1)^{-1}} &= \lim_{z\to c} \frac{\frac{1}{z}-\frac{1}{z-1}}{-1(z-1)^{-2}} \\
&= \lim_{z\to c} (z-1)^2 \bigg{(} \frac{1}{z-1}-\frac{1}{z}\bigg{)} \\
&= \lim_{z\to c} (z-1)^2 \frac{z-(z-1)}{z^2-z} \\
&= \lim_{z\to c} \frac{(z-1)^2}{z^2-z} \\
&= \lim_{z\to c} \frac{2(z-1)}{2z-1} \\
&= \lim_{z\to c} \frac{2z-2}{2z-1}
\end{align}
Similarly, the limit applied to the $Y_x^{(q)}$ and $X_y^{(q)}$ portions of the equation takes the form:
\begin{equation}
\lim_{z\to c} (z-1)log \bigg{(}\frac{z-1}{z}\bigg{)} = \lim_{z \to c} \frac{log(z-1)-log(z)}{(z-1)^{-1}}
\end{equation}
Then, applying l'Hopital's rule (and re-arranging, and applying l'Hopital again, etc):
\begin{align}
\lim_{z \to c} (z-1) log \bigg{(}\frac{z-1}{z}\bigg{)}
&= \lim_{z \to c} \frac{log(z-1)-log(z)}{(z-1)^{-1}} \\
&= \lim_{z \to c} \frac{\frac{1}{z-1}-\frac{1}{z}}{-1(z-1)^{-2}} \\
&= \lim_{z \to c} (z-1)^2 \bigg{(} \frac{1}{z} - \frac{1}{z-1} \bigg{)} \\
&= \lim_{z \to c} (z-1)^2 \frac{z-1-z}{z(z-1)} \\
&= \lim_{z \to c} \frac{-(z-1)^2}{z^2 - z} \\
&= \lim_{z \to c} \frac{-2(z-1)}{2z-1} \\
&= \lim_{z \to c} \frac{-(2z-2)}{2z-1}
\end{align}

This implies that for $z=N$ and $c = \infty$:
\begin{equation}
\lim_{N \to \infty} (N-1)log \bigg{(}\frac{N}{N-1}\bigg{)} = 1
\end{equation}
Although less intuitive, for $N \to \infty$, all relevant values of $A_{xy}^{(q)}$ will also approach infinity. This can be seen by revisiting the definition of $A_{xy}=N*p(x,y)$. We can then note that $p(x,y)$ is bounded between $0$ and $1$. This implies that for any value of $0<p(x,y)\leq 1$,  as $N \to \infty$, $A_{xy} \to \infty$. On the other hand, if $p(x,y)=0$, then $A_{xy}$ is also equal to zero. In this latter case, removing a $q$ data-point will {\it never} effect the bin ($A_{xy}^{(i \neq q)}={A'}_{xy}^{(i \neq q)}$), since that element of $A_{xy}$ contains no data-points; in other words, it will never exist in the limit in question. Based on this we can say that for $N \to \infty$:
\begin{equation}
\lim_{A_{xy}^{(q)} \to \infty} (A_{xy}^{(q)}-1)log \bigg{(}\frac{A_{xy}^{(q)}}{A_{xy}^{(q)}-1}\bigg{)} = 1
\end{equation}

Based on a similar argument as was made for the limit applied to $A_{xy}^{(q)}$ we also see that for $N \to \infty$ the values of $X_y^{(q)}$ and $Y_x^{(q)}$ will also increasingly grow and approach infinity implying:
\begin{equation}
\lim_{X_y^{(q)} \to \infty} (X_y^{(q)}-1)log \bigg{(}\frac{X_y^{(q)}-1}{X_y^{(q)}}\bigg{)} = -1, \text{  and  }
\lim_{Y_x^{(q)} \to \infty} (Y_x^{(q)}-1)log \bigg{(}\frac{Y_x^{(q)}-1}{Y_x^{(q)}}\bigg{)} = -1
\end{equation}

Finally, subbing the evaluations of these limits back into Equation \ref{limIq} we find that:
\begin{equation}
\lim_{N\to\infty} \frac{1}{N} \displaystyle\sum_i^{N} I(i) = I(x,y) + 1 - 1 - 1 + 1 = I(x,y)
\end{equation}
Which is, reassuringly, precisely the linear assumption made by the LIONESS equation. Thus, we note that for large values of $N$ Equation \ref{limIq} reduces to the original data-point version of mutual information defined in Equation \ref{MIi}.

We note that this evaluation relies on the {\it theoretical} limit of large $N$. However, as is often the case in mathematical biology, calculating MI in practice relies on real-world, finite data; we test this type of application explicitly in the main text of the manuscript. In the context of large, but finite, data one could postulate that the above limits might not always hold. This is true. However, we emphasize that correctly binning data is a recognized issue for correctly calculating {\it aggregate} MI-networks and is independent of whether one would then want to apply LIONESS to these network-models. The importance of binning data in a way that prevents the existence of bins with only a small number of data-points is well-recognized by scientists that use MI to estimate network models. Indeed, many computational approaches that compute MI bin data into only 3-4 bins to ensure that enough data-points are in each bin. 

\break
\subsection{Convergence properties of the LIONESS equation}\label{section_convergence}

Here we explore the behavior of the LIONESS equation in the limit of a large number of samples. For simplicity, let us define a new variable, $D=e_{ij}^{(\alpha)}-e_{ij}^{(\alpha-q)}$ and reiterate the LIONESS equation (Equation \ref{LIONESS}) in terms of $D$:
\begin{align}
e_{ij}^{(q)} &= N(e_{ij}^{(\alpha)}-e_{ij}^{(\alpha-q)})+e_{ij}^{(\alpha-q)}\\
&=ND+e_{ij}^{(\alpha-q)}
\end{align}
As the number of samples ($N$) approaches infinity, we note that $D$ approaches zero. In this limit the LIONESS equation can be thought of as a very large number times a very small number, plus a constant:
\begin{equation}
e_{ij}^{(q)}=[N\rightarrow \infty][D\rightarrow 0]+[e_{ij}^{(\alpha-q)}].
\end{equation}
This implies that, in the limit of large $N$, any sort of potential convergence behavior of the LIONESS equation is dependent on the relationship between $D$ and $N$.

We have systematically investigated the consequence of this relationship in the context of the analysis presented in the main text. To begin, we used the Pearson correlation to estimate aggregate networks using the {\it in silico} data. We determined $D$ by subtracting the values estimated for edges in an aggregate network reconstructed using $N$ samples with the values estimated for edges in an aggregate network based on those same $N$ samples minus one. We then took the absolute value of these differences and determined the maximum across all edges in the network. We repeated this $1000$ times using different sets of $N$ samples, and in Figure \ref{derivations_fig}A plot the mean, plus and minus the standard deviation, of the maximum-difference values across the $1000$ sample-sets, and for varying values of $N$ (blue line). For reference a line at $y=x/N$ is also plotted (dashed black line). We see that the difference between $e^{(\alpha)}$ and $e^{(\alpha-q)}$ consistently reaches values greater than $1/N$. This indicates that $ND$ does not converge to zero and the LIONESS equation does not converge to $e_{ij}^{(\alpha-q)}$.

In the main text we apply LIONESS to four different network reconstruction approaches: Pearson Correlation, PANDA, Mutual Information (MI) and Context Likelihood of Relatedness (CLR). For each of these approaches, we have determined the median and interquartile range for all values of $D$ used to calculate the $N=10000$ single-sample networks based on the {\it in silico} expression data-set. We report these values (times $10000$) in the subset table in Figure \ref{derivations_fig}A. Unsurprisingly, the median value for $ND$ for all the methods is very close to zero. In addition, all approaches have an $ND$ interquartile range that is non-zero, and typically varies between 0.01 and 1.

In examining these values, we feel is it important to point out that each of these four reconstruction approaches estimates scores for network edges in a different way. As a consequence, the distribution of predicted edge weights in the associated aggregate models is different for each approach; these differences are reflected in $D$. For example, when reconstructing a network, CLR performs a joint Z-score normalization on a Mutual Information Matrix (MIM). As a part of this normalization step, it removes any edge with a MI value below either its row or column's mean in the MIM, assigning it a weight of zero. This step ``removes'' 51.4\% of edges in the CLR {\it in silico} aggregate network (assigning them a predicted weight of zero). This can be observed by examining the distribution of edge weights predicted for the CLR aggregate network (Figure \ref{derivations_fig}B). Consistently, 51.4\% of $D$ values based on CLR aggregate networks are also zero, as evidenced by the quartiles shown in the inset Table in Figure \ref{derivations_fig}A.


      \counterwithin{figure}{section}
      \setcounter{figure}{0} 
      \renewcommand{\thefigure}{E\arabic{figure}} 
      \renewcommand{\figurename}{Figure}

\begin{figure}[!bh]
\centering
\subfigure[Relationship between $D$ and $N$ in Pearson Networks]{\includegraphics[width=250px]{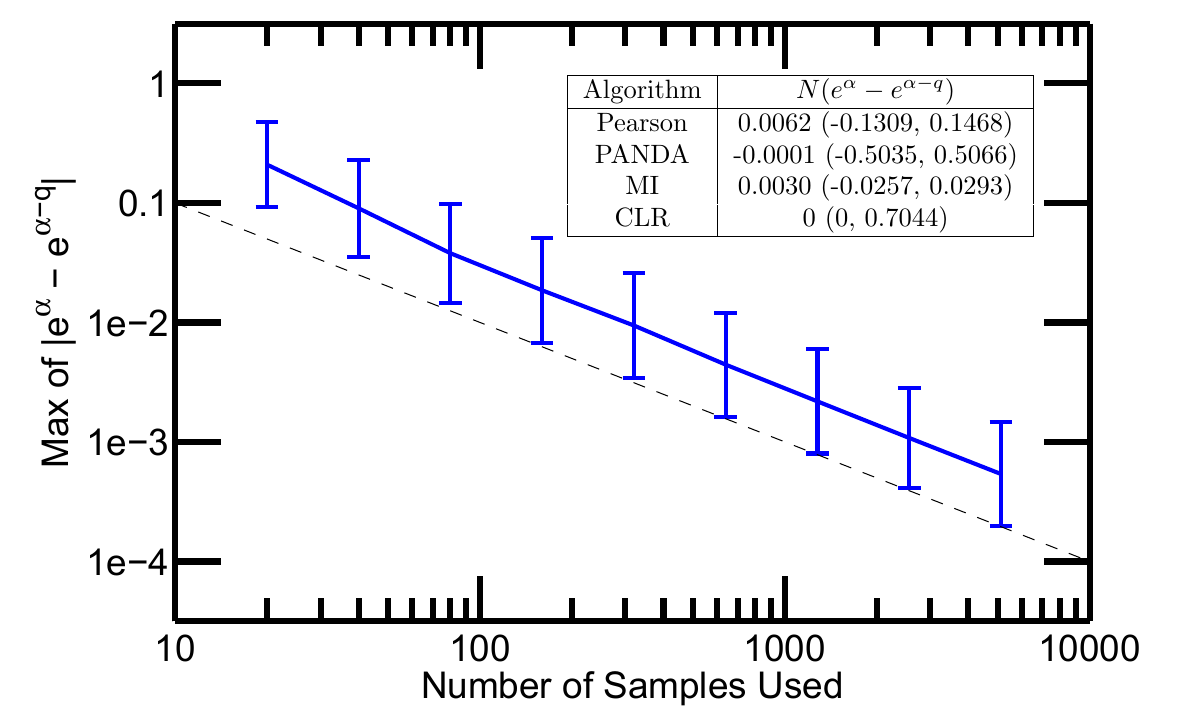}} 
\subfigure[CLR Aggregate Network]{\includegraphics[width=250px]{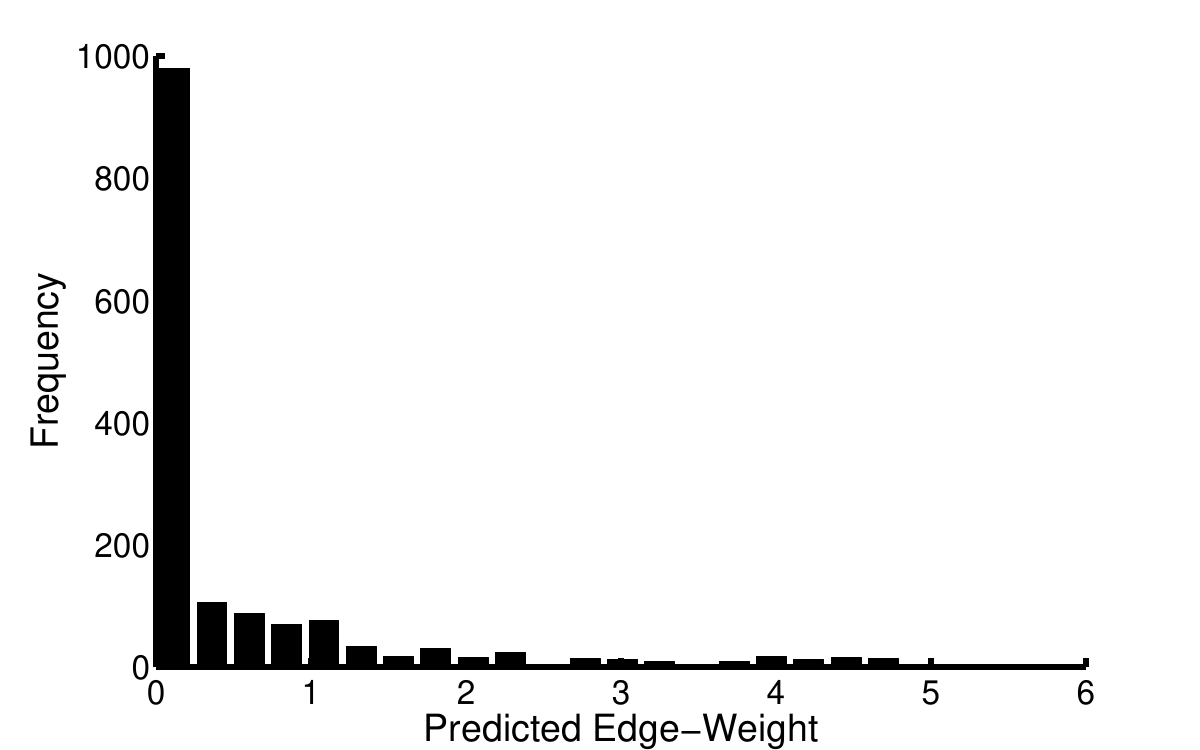}} 
\caption{(A) A plot of the maximum absolute value of $D$ across all edges when applying LIONESS to Pearson aggregate networks. Values were calculated for varying numbers of samples ($N$) and with different sets of samples. The mean plus/minus one standard deviation for these sets of samples is shown. The inset table shows the median and interquartile-range for the $ND$ values associated with the 10000 networks estimated using the four different reconstruction approaches explored in the main text.  (B) The distribution of the aggregate network ($e^{(\alpha)}$) edge-weights predicted by applying CLR to the {\it in silico} data. The values predicted by CLR are normalized Z-scores, but are always non-negative.}
\label{derivations_fig}
\end{figure}

\clearpage 

\end{landscape}
\end{document}